\documentclass[11pt]{article}
\pdfoutput=1 
\usepackage{amssymb}
\usepackage{amsmath}
\usepackage{amstext}
\usepackage{graphicx,epsfig}
\usepackage{epsfig}
\usepackage{verbatim} 
\usepackage{fancybox}
\usepackage{color}
\usepackage{ulem}
\usepackage{enumitem}
\usepackage{youngtab}
\usepackage{subfigure}
\usepackage{bbm}
\usepackage{parskip}
\usepackage[numbers,sort&compress]{natbib}
\usepackage{ytableau}
\usepackage[utf8]{inputenc}

\usepackage{tikz}
\usetikzlibrary{decorations}
\pgfdeclaredecoration{complete sines}{initial}
{
    \state{initial}[
        width=+0pt,
        next state=upsine,
        persistent precomputation={\pgfmathsetmacro\matchinglength{
            \pgfdecoratedinputsegmentlength / int(\pgfdecoratedinputsegmentlength/\pgfdecorationsegmentlength)}
            \setlength{\pgfdecorationsegmentlength}{\matchinglength pt}
        }] {}
    \state{upsine}[width=\pgfdecorationsegmentlength,next state=downsine]{
        \pgfpathsine{\pgfpoint{0.25\pgfdecorationsegmentlength}{0.5\pgfdecorationsegmentamplitude}}
        \pgfpathcosine{\pgfpoint{0.25\pgfdecorationsegmentlength}{-0.5\pgfdecorationsegmentamplitude}}
    }
    \state{downsine}[width=\pgfdecorationsegmentlength,next state=upsine]{
        \pgfpathsine{\pgfpoint{0.25\pgfdecorationsegmentlength}{-0.5\pgfdecorationsegmentamplitude}}
        \pgfpathcosine{\pgfpoint{0.25\pgfdecorationsegmentlength}{0.5\pgfdecorationsegmentamplitude}}
}
    \state{final}{}
}

\usepackage{genyoungtabtikz}
\Yboxdim{13pt} 
\Ylinethick{.8pt}

\newcommand*{\medoplus}{\raisebox{-0.3ex}{\scalebox{1.4}{$\oplus$}}}

\def\stu{{St\"uckelberg }}

\newcommand{\Comment}[1]{{}}
\definecolor{darkblue}{rgb}{0.15,0.35,0.55}
\definecolor{reddish}{rgb}{0.65, 0.2, 0.2}
\usepackage[linktocpage=true]{hyperref}
\hypersetup{
colorlinks=true,
citecolor=darkblue,
linkcolor=reddish,
urlcolor=darkblue,
pdfauthor={},
pdftitle={},
pdfsubject={}
}

\newcommand{\be}{\begin{equation}}
\newcommand{\ee}{\end{equation}}
\newcommand{\bea}{\begin{eqnarray}}
\newcommand{\eea}{\end{eqnarray}}
\newcommand{\beas}{\begin{eqnarray*}}
\newcommand{\eeas}{\end{eqnarray*}}
\newcommand{\nn}{\nonumber}

\def\mn{_{\mu \nu}}

\definecolor{darkred}{rgb}{0.7,0.3,0.3}
\definecolor{darkgreen}{rgb}{0.2,0.7,0.3}
\definecolor{lightgreen}{rgb}{.816,.94,.753}
\definecolor{greyish}{rgb}{.8,.8,.8}
\definecolor{darkblue2}{rgb}{0.3,0.4,0.9}
\usepackage{pifont}
\usepackage{xcolor,colortbl}

\def\({\left(}
\def\){\right)}

\newcommand{\rd}{{\rm d}}
\newcommand{\vp}{\varphi}

\newcommand{\ra}{\rangle}

\newcommand{\half}{\frac{1}{2}}

\def\gsim{ \lower .75ex \hbox{$\sim$} \llap{\raise .27ex \hbox{$>$}} }
\def\lsim{ \lower .75ex \hbox{$\sim$} \llap{\raise .27ex \hbox{$<$}} }

\def\xyma{\xymatrix@M.7em}
\def\xymas{\xymatrix@M.1em}
\newcommand{\ba}{\begin{eqnarray}}
\newcommand{\ea}{\end{eqnarray}}

\title{}
\author{}

\numberwithin{equation}{section}

\begin{document}
%
\renewcommand{\thefootnote}{\fnsymbol{footnote}}
~
\vspace{1.75truecm}
\begin{center}
{\huge \bf{Shift Symmetries in (Anti) de Sitter Space}}\\ \vspace{.2cm}
{\LARGE \bf{}}
\end{center} 

\vspace{1truecm}
\thispagestyle{empty}
\centerline{{\Large James Bonifacio,${}^{\rm a,}$\footnote{\href{mailto:james.bonifacio@case.edu}{\texttt{james.bonifacio@case.edu}}} Kurt Hinterbichler,${}^{\rm a,}$\footnote{\href{mailto:kurt.hinterbichler@case.edu}{\texttt{kurt.hinterbichler@case.edu}}} Austin Joyce,${}^{\rm b,}$\footnote{\href{mailto:austin.joyce@columbia.edu}{\texttt{austin.joyce@columbia.edu}}} and Rachel A. Rosen${}^{\rm b,}$\footnote{\href{mailto:rar2172@columbia.edu}{\texttt{rar2172@columbia.edu}}}}}
\vspace{.5cm}
 
\centerline{{\it ${}^{\rm a}$CERCA, Department of Physics,}}
 \centerline{{\it Case Western Reserve University, 10900 Euclid Ave, Cleveland, OH 44106}} 
 \vspace{.25cm}
 
 \centerline{{\it ${}^{\rm b}$Center for Theoretical Physics, Department of Physics,}}
 \centerline{{\it Columbia University, New York, NY 10027}} 
 \vspace{.25cm}

 \vspace{.8cm}
\begin{abstract}
\noindent
We construct a class of extended shift symmetries for fields of all integer spins in de Sitter (dS) and anti-de Sitter (AdS) space.  
These generalize the shift symmetry, galileon symmetry, and special galileon symmetry of massless scalars in flat space to all symmetric tensor fields in (A)dS space.
These symmetries are parametrized by generalized Killing tensors and exist for fields with particular discrete masses corresponding to the longitudinal modes of massive fields in partially massless limits.
We construct interactions for scalars that preserve these shift symmetries, including an extension of the special galileon to (A)dS space, and discuss possible generalizations to interacting massive higher-spin particles.
\end{abstract}

\newpage

\setcounter{tocdepth}{2}
\tableofcontents
\newpage
\renewcommand*{\thefootnote}{\arabic{footnote}}
\setcounter{footnote}{0}

\section{Introduction}
Shift symmetries play a powerful role in diverse areas of physics: they provide a useful classification of low-energy effective theories
and appear generically in any theory in which an internal or spacetime symmetry is spontaneously broken. In theories with spontaneously broken symmetries, masslessness of the Goldstone bosons is protected by symmetries that act like shift symmetries to leading order in powers of the fields.  The avatars of these symmetries in scattering amplitudes are enhanced soft limits, the prototypical example of which is the Adler zero \cite{Adler:1964um,Adler:1965ga}. Theories with shift symmetries are also known to enjoy various non-renormalization theorems \cite{Luty:2003vm,Hinterbichler:2010xn,deRham:2014wfa,Goon:2016ihr}.

The simplest example of a shift-symmetric theory is a free massless scalar field in flat space.  In fact, this theory has an infinite number of non-linearly realized symmetries which take the form of an infinite tower of shifts,\footnote{The free scalar also has an infinite number of unbroken symmetries, which form the higher-spin algebra underlying Vasiliev's higher-spin theory \cite{Eastwood:2002su,Vasiliev:2003ev,Bekaert:2005vh,Joung:2014qya}.  The shift symmetries studied here are distinct from these.}
\be
\delta\phi=c+c_\mu x^\mu+c_{\mu_1\mu_2}x^{\mu_1} x^{\mu_2}+c_{\mu_1\mu_2\mu_3} x^{\mu_1}x^{\mu_2}x^{\mu_3} +\cdots. \label{flatshiftgen}
\ee
Here the $c_{\mu_1\cdots\mu_k}$ are rank-$k$ symmetric and traceless constant tensors, and $x^\mu$ are the Cartesian spacetime coordinates. 
We call $k$ the {\it level} of the shift symmetry. Interactions generically break the shift symmetries~\eqref{flatshiftgen}; however, certain classes of interactions can preserve subsets of these symmetries.   The symmetries preserved by interaction terms therefore provide a useful organizing principle for classifying derivatively-coupled effective field theories (EFTs) in flat space~\cite{Hinterbichler:2014cwa,Griffin:2014bta,Cheung:2016drk,Padilla:2016mno,Bogers:2018zeg,Pajer:2018egx}.

The $k=0$ shift symmetry is the standard shift by a constant, $c$. Any interacting theory involving at least one derivative per field will preserve this symmetry, including ghost-free theories such as $P(X)$ theories.  If we allow for multiple interacting fields, there exist interesting field-dependent deformations of the constant shift symmetry such that the symmetry generators form a non-abelian algebra.  Field theories invariant under these deformed symmetries are of the non-linear $\sigma$-model type.  

The $k=1$ shift symmetry underlies the galileon~\cite{Luty:2003vm,Nicolis:2008in}.  Any interaction with at least two derivatives per field preserves this symmetry, but there is also a finite set of interactions, the galileon interactions, that have fewer than two derivatives per field and are ghost-free,
\be 
{\cal L}_n\sim \phi S_{n-1}(\partial\partial\phi) \,,\qquad\quad n=1,2,\cdots,D+1 \, ,
\ee
where $S_n$ are the symmetric polynomials defined in Appendix \ref{xtensorappendix}.
These can be understood as Wess--Zumino terms for the galileon symmetry \cite{Goon:2012dy} and, from the point of view of the $S$-matrix, as theories with enhanced soft limits~\cite{Cheung:2014dqa,Cheung:2016drk,Padilla:2016mno}.\footnote{Enhanced in this context means that scattering amplitudes vanish more quickly in the soft limit than would naively be expected from the number of derivatives per field.} In this case, there are again interesting non-abelian deformations of the symmetries involving field-dependent terms, but here the deformation can be achieved with a single scalar field. For example, both the Dirac--Born--Infeld (DBI) action and the action of the conformal dilaton possess non-abelian symmetries that start linear in the spacetime coordinates.

For $k=2$, any interaction with at least three derivatives per field preserves the symmetry, but the known interactions  invariant under the abelian symmetry all result in higher-derivative equations of motion.  However, there is a unique theory with second-order equations of motion that is invariant under the following deformed, non-abelian version of the symmetry:
\be
\delta\phi = c_{\mu\nu}x^{\mu} x^{\nu}+\frac{1}{\Lambda^6}c_{\mu\nu}
\partial^{\mu} \phi \partial^{\nu} \phi, \label{eq:sgsymmytry}
\ee 
where $\Lambda$ is the strong coupling scale of the theory.
This theory, known as the special galileon~\cite{Cheung:2014dqa,Hinterbichler:2015pqa,Cheung:2016drk,Novotny:2016jkh}, has fewer than two derivatives per field and is ghost-free. It is a particular galileon theory made from all of the interactions of even order in the fields, with fixed relative coefficients.  This theory has no free parameters, other than $\Lambda$, and its $S$-matrix has a soft limit that is even more enhanced than the regular galileons as a result of the symmetry \eqref{eq:sgsymmytry}.  In fact, the theory is completely fixed by requiring this enhanced soft limit~\cite{Cheung:2015ota,Bogers:2018zeg}.
 
For $k\geq 3$, there is no known way to have ghost-free interactions \cite{Hinterbichler:2014cwa,Griffin:2014bta}, and there are no on-shell constructible $S$-matrices with corresponding enhanced soft limits~\cite{Cheung:2016drk}.

In this work, we extend this classification of shift-symmetric EFTs to maximally symmetric curved spacetimes and to particles with nonzero spin. The de Sitter (dS) and anti-de Sitter (AdS) space analogues of the flat space shifts, such as \eqref{flatshiftgen}, are transformations that shift (A)dS fields by polynomials of ambient space coordinates.
In (A)dS space, in contrast to flat space, free fields with such shift symmetries have nonzero masses. For each non-negative integer $k$, there is a particular mass for which a given (A)dS field has the analogue of the level-$k$ flat space shift symmetry.  
For example, scalar fields with masses given by
\be
m_k^2 = -k(k+D-1)H^2
\label{eq:scalarmass}
\ee
have the (A)dS analogue of the flat space shift symmetry of level $k$. 

The discrete nature of the mass values \eqref{eq:scalarmass} is reminiscent of the phenomenon of partial masslessness in (A)dS space \cite{Deser:1983tm,Deser:1983mm,Higuchi:1986py,Brink:2000ag,Deser:2001pe,Deser:2001us,Deser:2001wx,Deser:2001xr,Zinoviev:2001dt,Skvortsov:2006at,Boulanger:2008up,Skvortsov:2009zu,Alkalaev:2011zv,Joung:2012rv,Basile:2016aen}, whereby a massive spin-$s$ field develops a gauge invariance for particular discrete values of its mass. These masses are labelled by an integer, $t$, called the partially massless (PM) depth, where $t \in \{0, \, \ldots, \, s-1\}$ and $t=s-1$ corresponds to a massless field.  In fact, there is a relationship between these two sets of discrete masses: if we start with a generic massive spin-$s$ field and send its mass to the $t=0$ PM value, then this massive field will break up into a PM field and a scalar with a shift symmetry of level $k=s-1$. The scalar shift symmetries then correspond to the PM reducibility parameters, {\it i.e.}, the global symmetry part of the PM gauge symmetry.

Scalar fields with the mass values~\eqref{eq:scalarmass} appear in a variety of contexts.  In the AdS case, these scalars have positive masses and are unitary. In dS space, although they are tachyonic, scalar fields with these particular mass values also belong to unitary representations~\cite{vilenkin1978special,Gazeau:2010mn}.\footnote{We give a physical argument for their unitarity in dS space in Section~\ref{sec:unitarity}. In terms of representation theory, these scalars belong to the so-called exceptional series~\cite{Basile:2016aen}. In CFT language, the conformal primary associated to one of these bulk tachyonic scalars should correspond to the boundary value of the shift-invariant ``curvature" obtained by taking $k+1$ derivatives of the bulk field. It would be interesting to further elucidate the representation theory of scalar fields belonging to the family~\eqref{eq:scalarmass}.}
Some aspects of their quantization are discussed in~\cite{Folacci:1992xc, Bros:2010wa, Epstein:2014jaa}, while other features are noted in~\cite{Shaynkman:2000ts, Chekmenev:2015kzf}. The $k=0$ case corresponds to the massless minimally coupled scalar and has been well studied (see, {\it e.g.},~\cite{Allen:1985ux,Allen:1987tz}), and the $k=1$ case has been studied as a toy model for the conformal mode of the graviton~\cite{Antoniadis:1991fa,Folacci:1996dv}. These scalar tachyons also appear in the context of CFT entanglement entropy~\cite{deBoer:2015kda, deBoer:2016pqk}, and scalars with negative integer conformal weights play an important role in the constructions of Ref.~\cite{Arkani-Hamed:2018kmz}. As mentioned in~\cite{Baumann:2017jvh} and discussed in Appendix~\ref{app:GJMSscalar}, these scalars are also connected to higher-order conformal scalars in even dimensions.

In fact, the shift-symmetric scalar fields fit into a larger picture: for particular discrete masses, free fields with nonzero spin in (A)dS space are also invariant under shifts by certain polynomials of ambient space coordinates.  The presence of these symmetries can similarly be traced to (higher-depth) PM fields.
As we explain, these fields are the longitudinal modes of massive fields as their masses approach the $t>0$ PM values, and the shift symmetries descend from PM reducibility parameters. 
We explain how this works for fields of all integer spins.

After classifying these shift symmetries for free bosonic fields in (A)dS space, we consider
interactions for (A)dS scalars that preserve the shift symmetries. The goal is to classify interesting (A)dS EFTs using shift symmetries as an organizing principle, in the same spirit as in flat space. There are interactions that are invariant under the undeformed symmetries of the free theory for any value of $k$, but only for $k=0$ and $k=1$ are there ghost-free interactions. The $k=0$ theories are $P(X)$ theories in (A)dS space, while the $k=1$ theory is the (A)dS galileon~\cite{Goon:2011uw,Goon:2011qf,Burrage:2011bt}.

For $k=1, \,2$, the algebra of symmetries underlying these scalar theories can be deformed in a unique way.  The $k=1$ deformed algebra is a real form of $\frak{so}(D+2)$.\footnote{Here and throughout, we write the complexified algebra without specifying the real form, since the particular real form depends on the spacetime signature, whether we are in dS or AdS space, and on the sign of certain parameters in the algebras, for example $\alpha$ in~\eqref{k1commutatoralge} and~\eqref{k2commutatoralge}. } Interacting theories are known from brane~\cite{Goon:2011uw,Goon:2011qf,Burrage:2011bt} and coset~\cite{Clark:2005ht,Hinterbichler:2012mv} constructions that realize symmetry breaking from $\frak{so}(D+2)$ down to the $\frak{so}(D+1)$ isometry algebra of (A)dS$_D$, and we describe the invariant interactions for a choice of field variables where the symmetry transformation takes a particular form in ambient space. For $k=2$, the deformed algebra is $\frak{sl}(D+1)$ and we find in all dimensions a nonlinear theory that realizes the breaking of this symmetry to the (A)dS isometry algebra. This theory is the analogue of the special galileon in (A)dS space. It involves a rather intricate structure of interactions that are fixed by the (A)dS version of the special galileon symmetry. For example, in $D=4$ it takes the form
\begin{align}
{1\over \sqrt{-g}}{\cal L}_{\rm SG}=&  -\frac{\Lambda ^6 }{ H^2}\frac{ (y^2-8 y+8) \left(8 X^2-3 y^{3/2} \sqrt{X+y}+12 X y-3 X \sqrt{y} \sqrt{X+y}+3 y^2\right)}{15y^3
   (X+y)^{3/2}} \nn\\ 
&   -\frac{\Lambda ^6 }{ H^2} \left(\frac{ 5 (y-4) y+16}{10y^{5/2}}-{1\over 10}\right)    
+\frac{2 (y-4) \phi }{15 X y^{5/2}} \left(\frac{\sqrt{y} (2 X+3 y)}{(X+y)^{3/2}}-3\right){H^2\over \Lambda ^6}\partial^\mu\phi\partial^\nu\phi X^{(1)}_{\mu\nu}(\Pi) \nn\\
&+\frac{y-2}{30  X^2 y^2} \left(2 \sqrt{y}-\frac{2 X^2+3 X y+2 y^2}{(X+y)^{3/2}}\right){1\over \Lambda ^6}\partial^\mu\phi\partial^\nu\phi X^{(2)}_{\mu\nu}(\Pi) \nn\\
&+\frac{\phi }{45 X^2 y^{3/2}} \left(\frac{\sqrt{y} (3 X+2 y)}{(X+y)^{3/2}}-2\right){H^2\over \Lambda^{12}}\partial^\mu\phi\partial^\nu\phi X^{(3)}_{\mu\nu}(\Pi),\nonumber
\end{align}
where
\be y\equiv 1+4{H^4\over \Lambda^6}\phi^2,\quad X\equiv {H^2\over \Lambda^6}(\partial\phi)^2, \quad \Pi_{\mu \nu} \equiv \nabla_{\mu } \nabla_{\nu} \phi \, ,\ee
and the tensors $X^{(j)}_{\mu\nu}$ are defined in Appendix~\ref{xtensorappendix}.
The highest-derivative terms are those of the flat space special galileon, but there are highly nontrivial lower-derivative terms suppressed by the (A)dS radius, including a potential, that are fixed by the symmetry. We expect that this theory should have similarly compelling properties to the special galileon in flat space.

We begin in Section~\ref{linesymmsectn} by describing certain shift symmetries enjoyed by free bosonic fields with particular discrete masses in maximally symmetric spacetimes. In Section~\ref{PMsection}, we explain the relation between these fields and PM fields. We consider interactions in Section~\ref{sec:interactingscalars}, focusing on the case of a single scalar field. We first classify possible deformations of the symmetry algebras of the free theories and then find interactions invariant under either the undeformed or deformed symmetries. We conclude in Section \ref{sec:conclusions} and discuss possible generalizations of our results, including interactions for massive higher-spin particles.  We include some technical results and reviews of useful material in the appendices:
the symmetric polynomials and the tensors $X_{\mu \nu}^{(n)}$ are defined in Appendix~\ref{xtensorappendix}, the ambient space formalism and some useful embedding coordinates are reviewed in Appendix~\ref{embbedappendix}, we describe the construction of scalar interactions using nonlinear realization techniques in Appendix~\ref{cosetappendix}, the relation between the shift-symmetric scalars and conformal powers of the Laplacian is discussed in Appendix~\ref{app:GJMSscalar}, and we discuss the PM decoupling limits of a massive spin-3 particle  in Appendix \ref{spin3Appendix}.

\vspace{.15cm}
\noindent
{\bf Conventions:}
We denote the spacetime dimension by $D$ and define $d \equiv D-1$. We use the mostly plus metric signature convention.    We denote the dS space Hubble scale as $H$, so that the Ricci scalar is $R=D(D-1)H^2>0$.   We denote the AdS space radius by $L$, so that $R=-{D(D-1)/ L^2}<0$.  We can go between the two cases with the relation $H^2\leftrightarrow {-1/L^2}$.   We sometimes use ${\cal R}$ for the radius to cover both cases at once, so that ${\cal R}=1/H$ for dS space and ${\cal R}=L$ for AdS space. Though we phrase things in Lorentzian signature, the theories we consider can be analytically continued to live on Euclidean spheres or hyperbolic spaces straightforwardly.  Tensors are symmetrized and antisymmetrized with unit weight, {\it e.g.}, $T_{(\mu\nu)}=\half \left(T_{\mu\nu}+T_{\nu\mu}\right)$,   $T_{[\mu\nu]}=\half \left(T_{\mu\nu}-T_{\nu\mu}\right)$, and $(\cdots)_T$ indicates that we take the symmetric fully traceless part of the enclosed indices, {\it e.g.}, $T_{(\mu\nu)_T}=\half \left(T_{\mu\nu}+T_{\nu\mu}\right)-{1\over D}g_{\mu\nu}T^\rho{}_{\rho}$. We denote Young projectors by ${\cal Y}_{r_1,r_2,r_3,\ldots}$, where the $r_i$ label the lengths of the rows of the corresponding Young diagrams. Our convention for the PM depth is that the depth, $t$, labels the number of indices of the PM gauge parameter.


\section{Shift symmetries of free fields in (A)dS space\label{linesymmsectn}}


We begin by describing how free bosonic fields in (A)dS space with particular discrete masses enjoy extended shift symmetries that are polynomials in the ambient space coordinates. 

\subsection{Scalar fields}
\label{sec:linearscalars}
First we consider scalar fields.
When we extend the free massless scalar action to (A)dS space, only the constant shift symmetry remains unbroken.
However, for each $k$ there is a particular mass for which the massive (A)dS action is invariant under the analogue of the level-$k$ flat space shift symmetry.

Consider the action
\be 
S=\int \rd^D x \, \sqrt{-g}\left(-\frac{1}{2}(\partial\phi)^2-\frac{m_k^2}{2} \phi^2\right) ,\label{massivegenadse}
\ee
where
\be 
m_k^2=-k(k+D-1)H^2\, .\label{scalarmassvaluese}
\ee
This action has a shift symmetry given by
\be
\delta \phi = K^{(k)},\label{adsksymmembe}
\ee
where $K^{(k)}$ is the restriction to (A)dS space of a degree-$k$ homogeneous polynomial of ambient space coordinates,
\be
K^{(k)}  = S_{A_1\cdots A_k}X^{A_1}\cdots X^{A_k}\big|_{\rho = \mathcal{R}} \, .
\ee
Here $X^{A}(x)$ is an embedding of (A)dS$_D$ into a $(D+1)$-dimensional flat ambient space, restriction to the (A)dS hyperboloid is denoted by $|_{\rho = \mathcal{R}}$, and $S_{A_1\cdots A_k}$ is a constant ambient space tensor that is symmetric and traceless (for more details on the ambient space formalism, see Appendix \ref{embbedappendix}). 
Associated with $\phi$ is an ambient space scalar $\Phi$ that has homogeneity degree $k$ and equals $\phi$ on the (A)dS surface,
\be
\left(X^A\partial_A - k\right)\Phi = 0\,,~~~~~~~~~~~~~~~~~\phi(x) = \Phi(\rho, x) \big|_{\rho = \mathcal{R}} \,,
\ee 
so the shift symmetry \eqref{adsksymmembe} corresponds to the ambient space transformation
\be \label{eq:scalarambientshift}
\delta \Phi  = S_{A_1\cdots A_k}X^{A_1}\cdots X^{A_k}.
\ee
The number of independent components of $S_{A_1\cdots A_k}$ is 
\be
N_{\rm symm.}=\left( \begin{matrix} D+k \\ k \end{matrix} \right)-\left( \begin{matrix} D+k-2\\ k-2 \end{matrix} \right) \, ,
\label{eq:numcomponsymmtracetensor}
\ee
so there are this many independent symmetries for a given $k$.

Unlike the massless scalar in flat space, which has a symmetry for each value of $k$, the (A)dS action \eqref{massivegenadse} has the symmetry only for a single value of $k$. Placing the scalar theory on curved spacetime therefore splits the infinite number of shift symmetries of the flat space theory.
Conversely, in the flat limit the (A)dS shift symmetry of level $k$ becomes the flat space symmetries with levels $\leq k$.
For example, the $k=1$ (A)dS symmetry has $D+1$ components, $S_A$, which in the flat limit become the $D$ galileon symmetries, $c_\mu$, and the shift symmetry, $c$.  The $k=2$ (A)dS symmetry is described by the $D(D+3)/2$ components of the traceless symmetric tensor, $S_{AB}$. In the flat limit, these become the $(D+2)(D-1)/2$ linear special galileon symmetries, $c_{\mu\nu}$, the $D$ galileon symmetries, $c_\mu$, and the shift symmetry, $c$. In the general case, the splitting is described by the following branching rule:
\be
\gyoung(_4k)^{\,T}\xrightarrow[{\scriptscriptstyle D+1\rightarrow D}]{}
~\gyoung(_4k)^{\,T}\,
\raisebox{0.235ex}{\medoplus}~\,
\gyoung(_3{k-1})^{\,T}\,
\raisebox{0.235ex}{\medoplus}~\,
\cdots\,
\raisebox{0.235ex}{\medoplus}~\,
\gyoung(;)~\,
\raisebox{0.235ex}{\medoplus}~\,
\bullet\,.
\ee

It is straightforward to verify that the free scalar action \eqref{massivegenadse} has the symmetry \eqref{adsksymmembe}.  Due to the tracelessness of the ambient space tensor $S_{A_1\cdots A_k}$, the ambient space transformation \eqref{eq:scalarambientshift} satisfies
\be
\partial_A \partial^A \delta\Phi=0.
\ee
Pulled back to the (A)dS space, this becomes
\be
\left( \square +k(k+D-1)H^2\right)\delta\phi=0,
\ee 
which is precisely the Klein--Gordon equation derived from the action \eqref{massivegenadse}. Since the free action is quadratic in the field, any solution to the equation of motion is a symmetry of the action.\footnote{The symmetry transformations in \eqref{adsksymmembe} are precisely the spherical harmonics on (A)dS space, which are eigenfunctions of the (A)dS Laplacian with eigenvalues equal to minus the mass values \eqref{scalarmassvaluese}. We could also consider shifts by other solutions to the free equations of motion. However, as we will see, the spherical harmonics are distinguished by their connection to PM reducibility parameters and lead to interesting invariant interactions. 
}

As discussed in the introduction, the scalar fields described by \eqref{massivegenadse} appear in myriad theoretical contexts. Despite this, they are tachyonic in dS space and thus one might doubt whether they can be physical there. However, the naively worrisome growing modes of these fields can be removed precisely by the shift symmetries discussed in this section. To see this, recall that in the inflationary slicing of dS space the zero mode of a scalar field of mass $m$ evolves at late times as
\be
\phi_{\vec q=0}(\eta) \sim \alpha\eta^{\Delta_{+}}+\beta\eta^{\Delta_{-}},
\label{eq:dSgrowingmode}
\ee
where $\Delta_{\pm}$ are the two fall-offs defined  in terms of the mass through the relation
\be
\Delta_{\pm} = \frac{d}{2}\pm \sqrt{\frac{d^2}{4}-\frac{m^2}{H^2}}.
\ee
For the mass values~\eqref{scalarmassvaluese}, the two fall-offs correspond to
\be
\Delta_-  = -k,~~~~~~~~~~~~~~\Delta_+  = d+k.
\ee
The fact that $\Delta_-$ is negative is a symptom of the tachyonic mass---at late times ($\eta\to 0$) the field is diverging. However, this dangerous zero mode can be removed by a shift transformation. To see this, we note that the inflationary slicing is given by the embedding~\eqref{eq:dsflatembedding}, so that if we perform a shift~\eqref{adsksymmembe} along the lightcone coordinate~\eqref{eq:xpluslightcone}, we find
\be
\delta\phi = S_{+\cdots +}X^{+}\cdots X^{+}\big|_{\rho = \mathcal{R}} \propto  \eta^{-k},
\ee
which has precisely the same time dependence as the growing mode~\eqref{eq:dSgrowingmode}. This is not completely satisfactory, because modes with arbitrarily small wave numbers will also seem unstable if we wait long enough. However, this is an artifact of the inflationary slicing. Going to global coordinates, there are only a finite number of modes that are sick: modes with spatial angular momentum $L\leq k$ are tachyonic.\footnote{More properly, they are zero-norm with respect to the Klein--Gordon inner product.} However, these are precisely the modes which can be generated or removed by the symmetries~\eqref{scalarmassvaluese}, so that these scalar fields appear to be healthy.

In the AdS context, these scalars are not tachyonic, but rather correspond to conventional scalar representations.  They lie well above the Breitenlohner--Freedman bound~\cite{Breitenlohner:1982jf} and are therefore unitary.  These particular mass values and their associated shift symmetries have not, to our knowledge, been much studied in AdS space, though we expect that they should play some interesting role in the AdS/CFT correspondence.

\subsection{Symmetric tensor fields}

Shift symmetries for fields with spin $s\geq 1$ have not been extensively studied, likely because Goldstone bosons for broken internal symmetries are always scalars.\footnote{Additionally, massless Goldstone fields with nonzero spin coming from broken spacetime symmetries are subject to constraints, see {\it e.g.},~\cite{Klein:2018ylk}.} However, in certain cases massive Goldstone-like fields with spin arise in (A)dS space \cite{deRham:2018svs} (soft limits for spin-1 theories in flat space have been studied in \cite{Cheung:2018oki}). We now discuss how the scalar shift symmetries generalize to symmetric tensor fields in (A)dS space.

A free spin-$s$ bosonic field of mass $m$ on ${\rm (A)dS}_D$ can be described by a completely symmetric tensor, $\phi_{\mu_1\cdots\mu_s}$, that obeys the following on-shell equations of motion:
\be
\left(\square-H^2\left[s+D-2-(s-1)(s+D-4)\right]-m^2\right)\phi_{\mu_1\cdots\mu_s} = 0\,,\quad \nabla^\nu\phi_{\nu\mu_2\cdots\mu_s} = 0\,,\quad\phi^\nu_{\ \nu\mu_3\cdots\mu_{s}} = 0 \,.
\label{massivefields}
\ee
This theory develops a shift symmetry at the following values of the mass:
\be
m^2_{s,k}= -(k+2) (k+D-3+2 s)H^2 \,, ~~\qquad k=0,1,2,\ldots,\quad  (s\geq 1)\,. \label{spinsmassvalueseqe}
\ee
The form of the level-$k$ shift symmetry acting on the spin-$s$ field is 
\be \label{eq:tensorshift}
\delta \phi_{\mu_1\cdots\mu_s}=K_{\mu_1\cdots \mu_s}^{(k)} \, ,
\ee
where the tensor on the right-hand side is given by
\be 
K_{\mu_1\cdots \mu_s}^{(k)} = S_{A_1\cdots A_{s+k},B_1\cdots B_s}X^{A_1}\cdots X^{A_{s+k}} {\partial X^{B_1}\over \partial x^{\mu_1}}\cdots {\partial X^{B_s}\over \partial x^{\mu_s}}\bigg|_{\rho = \mathcal{R}} \,.\label{Ktensordef1e}
\ee
The ambient space tensor $S_{A_1\cdots A_{s+k},B_1\cdots B_s}$ is a fully traceless, constant tensor with the symmetries of the following Young tableau:
%
%
\be 
S_{A_1\cdots A_{s+k},B_1\cdots B_s}\in ~\raisebox{1.15ex}{\gyoung(_5{s+k},_3s)}^{\,T} \, .
\label{repKexpre}
\ee
Associated with $\phi_{\mu_1\cdots\mu_s}$ is an ambient space tensor, $\Phi_{A_1\cdots A_s}$, with homogeneity degree $s+k$, that projects to $\phi_{\mu_1\cdots\mu_s}$ on the (A)dS surface,
\be
\phi_{\mu_1\cdots\mu_s}(x) = \Phi_{A_1\cdots A_s}(\rho, x)  {\partial X^{A_1}\over \partial x^{\mu_1}}\cdots {\partial X^{A_s}\over \partial x^{\mu_s}}\bigg|_{\rho = \mathcal{R}} \,,
\ee
so the shift symmetry \eqref{eq:tensorshift} corresponds to the ambient space transformation
\be
\delta \Phi_{B_1\cdots B_s} = S_{A_1\cdots A_{s+k},B_1\cdots B_s}X^{A_1}\cdots X^{A_{s+k}}.
\ee

The tensors \eqref{Ktensordef1e} are the spin-$s$ transverse-traceless spherical harmonics on (A)dS space. They also correspond to generalized traceless Killing tensors, which are the reducibility parameters for PM gauge transformations, as discussed in Section~\ref{sec:killingtensors}. They solve the free equations of motion \eqref{massivefields} with masses \eqref{spinsmassvalueseqe}, so shifting by these preserves the massive equations of motion and the free action.   
The massive spin-$s$ action for $s\geq 3$ also necessarily contains auxiliary fields that vanish on shell, but these do not transform under the shift symmetry.

\subsection{Examples}
To be concrete, let us write down a few explicit free actions and their corresponding shift symmetries. We focus on fields of spin $s=0,1,2$, with $k=0$. These examples illustrate how the $k=0$ shifts correspond to traceless Killing tensors in (A)dS space, as discussed further in Section~\ref{sec:killingtensors}. 

For $s=0$ and $k=0$, the free action is just that of a massless scalar field,
\be 
S=-\frac{1}{2}\int \rd^D x \, \sqrt{-g}(\partial \phi)^2\, .
\ee
The shift symmetry is a constant both in ambient space and in (A)dS space, \textit{i.e.},
\be
\delta \phi=\delta \Phi = K^{(0)}\,,
\ee
where $\partial_\mu K^{(0)} =0$.

For $s=1$ and $k=0$, the free action is that of a Proca field, $\phi_\mu$, of mass $m^2 = -2(D-1) H^2$,
\be 
S=\int \rd^D x \, \sqrt{-g}\left(-\frac{1}{4}(\nabla_\mu\phi_\nu-\nabla_\nu\phi_\mu)^2+(D-1) H^2\phi_\mu \phi^\mu\right)\, .
\ee
The transformation of the ambient space vector is given by
\be
\delta \Phi_B = S_{AB}X^A \, ,
\ee
where $S_{AB}$ is a constant antisymmetric tensor. This projects to the (A)dS space transformation
\be
\delta \phi_\mu = K^{(0)}_\mu \, ,
\ee
where $K^{(0)}_{\mu}$ is an (A)dS Killing vector,
\be
\nabla_\mu K^{(0)}_{\nu}+\nabla_\nu K^{(0)}_{\mu} =0 \,.
\ee  
There are $D(D+1)/2$ such vectors, in agreement with the number of components of $S_{AB}$.

For $s=2$ and $k=0$, the free action is that of a Fierz--Pauli massive graviton, $\phi_{\mu\nu}$, with mass $m^2 = -2(D+1) H^2$,
\begin{align}
S=\int \rd^D x \, \sqrt{-g} & \left(-\frac{1}{2}\nabla_\alpha \phi_{\mu\nu}\nabla^\alpha \phi^{\mu\nu}+\nabla_\alpha \phi_{\mu\nu}\nabla^\nu \phi^{\mu\alpha}-\nabla_\mu \phi^\alpha_{~\alpha} \nabla_\nu \phi^{\mu\nu}+\frac{1}{2}\nabla_\mu\phi^\alpha_{~\alpha}\nabla^\mu\phi^\nu_{~\nu} \right. \nonumber \\
&~~\left.+(D-1) H^2\left(\phi^{\mu\nu}\phi_{\mu\nu} -\frac{1}{2}\phi^{\mu}_{~\mu}\phi^{\nu}_{~\nu}\right) +(D+1)H^2(\phi^{\mu\nu}\phi_{\mu\nu} -\phi^{\mu}_{~\mu}\phi^{\nu}_{~\nu})\right)\, .
\end{align}
The ambient space transformation is given by
\be
\delta \Phi_{B_1B_2} = S_{A_1 A_2, B_1 B_2} X^{A_1}X^{A_2}\, ,
\ee
where $S_{A_1 A_2, B_1 B_2}$ is a fully traceless, constant tensor with the symmetries of the following Young tableau:
\be
S_{A_1 A_2, B_1 B_2} \in\,\raisebox{1.175ex}{\gyoung(;;,;;)}^{ \,\, T}.
\ee
This projects to the (A)dS space transformation
\be
\delta \phi_{\mu\nu} = K^{(0)}_{\mu\nu} \, ,
\ee
where $K^{(0)}_{\mu\nu}$ is a traceless (A)dS Killing tensor,
\be
\nabla_\mu K^{(0)}_{\nu\lambda}+\nabla_\nu K^{(0)}_{\lambda\mu}+\nabla_\lambda K^{(0)}_{\mu \nu} =0, \quad K_\mu^{(0) \,\mu} = 0. \label{eq:Killingexample}
\ee
The number of solutions to \eqref{eq:Killingexample} is \cite{doi:10.1063/1.523488, doi:10.1063/1.527288}
\be
N_{KT} = \frac{(D+1)(D+2)(D+3)(D-2)}{12},
\ee
in agreement with the number of independent components of $S_{A_1 A_2, B_1 B_2}$.

\subsection{Algebra of linearized symmetries}
We have seen that for fields of any integer spin in (A)dS space, there are special values of the mass for which they develop shift symmetries that are polynomials of the ambient space coordinates. These are spacetime symmetries, so we would like to explore how these symmetries interact with the (A)dS isometries. In particular, we would like to compute the algebra of symmetries, involving both the extended shifts and spacetime Killing symmetries. We do this with an eye towards constructing deformations of these symmetry algebras.

The (A)dS isometries acting on the ambient space field $\Phi_{A_1\cdots A_s}$ take the form
\be
\delta_{J_{AB}}\Phi_{A_1\cdots A_s}\equiv J_{AB}\Phi_{A_1\cdots A_s}=\left(X_A\partial_B-X_B\partial_A\right)\Phi_{A_1\cdots A_s}+\sum_{i=1}^s \left({\cal J}_{AB}\right)_{A_i}^{\ C}\Phi_{A_1\ldots A_{i-1} C A_{i+1} \ldots A_s},
\label{eq:adssymmtrans}
\ee
where $\left({\cal J}_{AB}\right)_{C}{}^{D}\equiv\eta_{AC}\delta_B{}^{D}-\eta_{BC}\delta_A{}^{D}$ is the Lorentz generator in the vector representation. 
The isometries satisfy the commutation relations of the algebra $\mathfrak{so}(D+1)$,
\be \left[ J_{AB},J_{CD}\right]= \eta_{AC}J_{BD}-\eta_{BC}J_{AD}+\eta_{BD}J_{AC}-\eta_{AD}J_{BC}.\ee

The level-$k$ shift symmetry takes the following form in the ambient space:
\be \delta_{S^{A_1\cdots A_{s+k},B_1\cdots B_s}} \Phi_{C_1\cdots C_s} \equiv S^{A_1\cdots A_{s+k},B_1\cdots B_s} \Phi_{C_1\cdots C_s}={\cal Y}^{(T)}_{s+k,s}\left[ X^{A_1}\cdots X^{A_{s+k}} \delta^{B_1}_{(C_1}\cdots \delta^{B_s}_{C_s)_T}\right],
\ee
where ${\cal Y}^{(T)}_{s+k,s}$ is the Young projector  onto the traceless tableau \eqref{repKexpre}, which acts on the $A_i,B_i$ indices.
These symmetries are independent of the fields, so they have trivial commutators among themselves,
\be \left[S_{A_1\dots A_{s+k},B_1\dots B_s}, \, S_{C_1\cdots C_{s+k},D_1\cdots D_s}\right]=0\,.\label{linegenscommutee}\ee
The commutators of the shift generators with the (A)dS isometries are
\be \left[J_{BC}, S_{A_1\cdots A_{2s+k}}\right] = \sum_{i=1}^{2s+k} \left( \eta_{BA_i}S_{A_1 \dots A_{i-1} C A_{i+1}\dots A_{2s+k}}-\eta_{CA_i}S_{A_1 \dots A_{i-1} B A_{i+1}\dots A_{2s+k}} \right),\\
\ee 
which shows that they transform as tensors in (A)dS space.
The total symmetry algebra is thus the semi-direct product of the (A)dS algebra with the abelian algebra of its mixed-symmetry traceless tensor representation of type \eqref{repKexpre}. An interesting question is whether this algebra can be deformed so that the shift generators no longer commute.
In Section \ref{algebrassection}, we study possible deformations of this algebra in the scalar case, $s=0$.

\section{Shift symmetries from partially massless fields\label{PMsection}}
Although it is clear that the free theories with particular discrete masses described in Section~\ref{linesymmsectn} possess polynomial shift symmetries, the underlying origin and importance of these symmetries may be somewhat obscure. Our goal in this section is to elucidate how these symmetries are intimately connected to the phenomenon of partial masslessness in (A)dS space.
As explained below, the shift-symmetric fields are the longitudinal modes of massive fields when we take various PM limits, with the shift symmetries arising from the reducibility parameters of the PM gauge transformations.

\subsection{Partially massless fields}
We first review some facts about PM fields.
In contrast to flat space, on (A)dS space there are representations that are neither strictly massive or massless.
These PM fields are fields with particular values of the mass, $m$, for which the free action develops a gauge invariance~\cite{Deser:1983tm,Deser:1983mm,Higuchi:1986py,Brink:2000ag,Deser:2001pe,Deser:2001us,Deser:2001wx,Deser:2001xr,Zinoviev:2001dt,Skvortsov:2006at,Boulanger:2008up,Skvortsov:2009zu,Alkalaev:2011zv,Joung:2012rv,Basile:2016aen}.

A spin-$s$ field has $s$ PM points, which are labeled by an integer, $t$, called the PM {\it depth}, $t\in \left\{0,1,\ldots, s-1\right\}$. The mass of a spin-$s$ depth-$t$ PM field is
\be
\bar{m}^2_{s,t} = (s-t-1)(s+t+D-4)H^2\,,
\label{pmpoints}
\ee
so a massless field corresponds to $t = s-1$.
A depth-$t$ PM field possesses a gauge invariance with a rank-$t$ totally symmetric gauge parameter.  This gauge invariance removes the helicity $0, \, \pm1, \ldots, \, \pm t$ components of the massive field, leaving a field which propagates only the $\pm(t+1),\dots, \, \pm s$ helicities.

Combining Eqs.~\eqref{massivefields} and~\eqref{pmpoints}, the on-shell equations of motion for a PM field of spin-$s$ and depth-$t$ are~\cite{Deser:2001xr,Zinoviev:2001dt,Hallowell:2005np},
\be
\left(\square-H^2\left[D+s-2-t(D+t-3)\right]\right)\phi_{\mu_1\cdots\mu_s} = 0, \qquad \nabla^\nu\phi_{\nu\mu_2\cdots\mu_s} = 0, \qquad\phi^\nu_{\ \nu\mu_3\cdots\mu_{s}} = 0.
\label{spinsdepthteom}
\ee
The gauge transformation of the PM field is given by
\be
 \delta \phi_{\mu_1\cdots\mu_s} = \nabla_{(\mu_{t+1}}\nabla_{\mu_{t+2}}\cdots\nabla_{\mu_{s}}\xi_{\mu_{1}\cdots\mu_{t})}+\ldots \label{pmintogt},
\ee
where the ellipses stand for ${\cal O}(H^2)$ terms with fewer derivatives. Explicitly, we can write the full transformation in the following factorized form~\cite{Hinterbichler:2016fgl}:
\be
 \delta \phi_{\mu_1\cdots\mu_s} = 
\left\{\begin{array}{l}
{\cal Y}_{s}\left(\prod_{n=1}^{\frac{s-t}{2}}\left[\nabla_{\mu_n}\nabla_{\mu_{n+\frac{s-t}{2}}}+(2n-1)^2H^2g_{\mu_n\mu_{n+\frac{s-t}{2}}}\right]\right)\xi_{\mu_{s-t+1}\cdots\mu_s},~~~~~~~~~~\,{\rm for}~(s-t)~{\rm even},\ \ \vspace{.15cm}\\
{\cal Y}_{s}\left(\prod_{n=1}^{\frac{s-t-1}{2}}\left[\nabla_{\mu_n}\nabla_{\mu_{n+\frac{s-t-1}{2}}}+(2n)^2H^2g_{\mu_n\mu_{n+\frac{s-t-1}{2}}}\right]\right)\nabla_{\mu_{s-t}}\xi_{\mu_{s-t+1}\cdots\mu_s},~{\rm for}~(s-t)~{\rm odd},
\end{array}\right. \label{pmintogt2}
\ee
where ${\cal Y}_s$ is the Young projector onto the totally symmetric part. 
The gauge parameter, $\xi_{\mu_{1}\cdots\mu_{t}}$, is a totally symmetric tensor, which is itself restricted to satisfy the on-shell equations
\be 
\Big(\square  + H^2\left[(s-1)(D+s-2)-t\right]\Big)\xi_{\mu_{1}\cdots\mu_{t}}=0,~~\qquad \nabla^{\nu}\xi_{\nu\mu_{2}\cdots\mu_{t}}=0,~~\qquad \xi^\nu_{\ \nu\mu_{3}\cdots\mu_{t}} =0.\label{genresga}\ee

\subsection{Dual operators}

Through the AdS/CFT correspondence, a spin-$s$ field in ${\rm AdS}_D$ corresponds to a spin-$s$ primary operator in a ${\rm CFT}_{d}$ with $d=D-1$.  The mass of the field and the scaling dimension, $\Delta$, of the conformal primary are related by
\be m^2L^2=\begin{cases} \Delta(\Delta-d), & s=0\, ,\\ \left(\Delta+s-2\right)\left(\Delta-s-d+2\right), &  s\geq 1\, .  \end{cases}   \label{AdSCFTformpe} \ee
For a given mass, there are two ways to quantize the field in AdS space (that is, assign it a dual operator).   These correspond to the greater and smaller roots of \eqref{AdSCFTformpe}, $\Delta_{\pm}$:  $\Delta_+$ corresponds to the ``standard quantization,'' which covers the primaries with $\Delta>d/2$, and $\Delta_-$ corresponds to the ``alternate quantization" \cite{Klebanov:1999tb}, which covers the primaries with $\Delta<d/2$.  

The unitarity bound appropriate for AdS primary fields is \cite{Mack:1975je,Jantzen1977,Minwalla:1997ka}
\be \Delta\geq  \begin{cases}  {d-2\over 2}, & s=0\,, \\ s+d-2, & s\geq 1\, .\end{cases}\label{unitboundd}\ee
The standard quantization of the PM masses \eqref{pmpoints} corresponds to the conformal dimensions
\be \Delta_{s,t}=t+d-1\, . \label{confdimbse} \ee
These violate the bound \eqref{unitboundd}, except for the massless cases, $t=s-1$, which saturate it.\footnote{This implies that PM fields are {\it not} unitary in AdS space. Conversely, in dS space, the reality condition on the conformal algebra is different, and PM fields {\it are} unitary.}

A primary state $|\Delta\ra^{i_1\cdots i_s}$ with the conformal dimension \eqref{confdimbse} gives rise to a short multiplet of the conformal group \cite{Dolan:2001ih}.  In particular, there is a null descendant at the $(s-t)^{\rm th}$ level, of the form
\be P_{i_1}\ldots P_{i_{s-t}}  |\Delta\ra^{i_1\cdots i_s}=0\, .\label{bosonmultiplyconsce}\ee
In the language of Ref.~\cite{Penedones:2015aga}, this is called a type II shortening.  It is precisely these null states that are dual to fields with shift symmetries in the AdS space.

\subsection{Generalized Killing tensors}
\label{sec:killingtensors}

The shift symmetries of the fields dual to the PM null states descend from the PM reducibility parameters, which are choices of the gauge parameter for which the gauge transformation vanishes \cite{Barnich:2001jy}. 
The reducibility parameters for a spin-$s$ and depth-$t$ PM field are generalized Killing tensors of rank $t$ that satisfy an equation with $s-t$ derivatives~\cite{Skvortsov:2006at}. We denote these by $K_{\mu_1\cdots \mu_t}^{(k)}$, where $k= s-t-1$, in anticipation of their connection with the spinning (A)dS spherical harmonics from Section~\ref{linesymmsectn}. 

Given the gauge transformation laws \eqref{pmintogt2}, the reducibility parameters of a depth-$t$ spin-$s$ PM field must satisfy an equation of the schematic form
\be
\nabla_{(\mu_{t+1}}\nabla_{\mu_{t+2}}\cdots\nabla_{\mu_{t+k+1}}K^{(k)}_{\mu_{1}\cdots\mu_{t})}+\cdots =0,
\label{pmintogtred}
 \ee
 where $+\cdots$ stands for terms with fewer derivatives and corresponding factors of $H$. Explicitly, the equation satisfied by the reducibility parameters is
\begin{align}
\begin{array}{l}
{\cal Y}_{k+t+1}\prod_{n=1}^{\frac{k+1}{2}}\left[\nabla_{\mu_n}\nabla_{\mu_{n+\frac{k+1}{2}}}+(2n-1)^2H^2g_{\mu_n\mu_{n+\frac{k+1}{2}}}\right]K^{(k)}_{\mu_{k+2}\cdots\mu_{k+t+1}}=0,~~~~\,{\rm for}~k~{\rm odd},\ \ \vspace{.15cm}\\
{\cal Y}_{k+t+1}\prod_{n=1}^{\frac{k}{2}}\left[\nabla_{\mu_n}\nabla_{\mu_{n+\frac{k}{2}}}+(2n)^2H^2g_{\mu_n\mu_{n+\frac{k}{2}}}\right]\nabla_{\mu_{k+1}}K^{(k)}_{\mu_{k+2}\cdots\mu_{k+t+1}}=0,~~~~~~~{\rm for}~k~{\rm even}, \ \ 
\end{array}
\label{pmintogtred2}
\end{align}
where the symmetric Young projector ${\cal Y}_{k+t+1}$ acts on everything to its right.
In addition to~\eqref{pmintogtred2},
 from the on-shell conditions \eqref{genresga} we find that the reducibility parameters must satisfy
\be 
\Big(\square  + H^2\left[(k+t) (k+t+D-1)-t\right]\Big)K^{(k)}_{\mu_{1}\dots\mu_{t}}=0,\quad \nabla^{\nu}K^{(k)}_{\nu\mu_{2}\dots\mu_{t}}=0,\quad  K^{(k)\nu}{}_{\nu\mu_{3}\dots\mu_{t}} =0.\label{genresgared}\ee
The space of solutions to Eqs.~\eqref{pmintogtred} and \eqref{genresgared} is finite dimensional.  It is parametrized by precisely the tensors of the type \eqref{Ktensordef1e},
\be
 K_{\mu_1\cdots \mu_t}^{(k)}=K_{A_1\cdots A_{t+k},B_1\cdots B_t}X^{A_1}\cdots X^{A_{t+k}} {\partial X^{B_1}\over \partial x^{\mu_1}}\cdots {\partial X^{B_t}\over \partial x^{\mu_t}}, \label{eq:killingkt}
\ee
where $X^{A}(x)$ are ambient space coordinates and $K_{A_1\cdots A_{t+k},B_1\cdots B_t}$ is a fully traceless, constant tensor with the symmetries of the Young tableau
%
\be 
K_{A_1\cdots A_{t+k},B_1\cdots B_t}\in\, 
\raisebox{1.15ex}{\gyoung(_5{t+k},_3t)}^{\,T}.
\ee

For $k=0$, the tensors \eqref{eq:killingkt} are ordinary traceless rank-$t$ Killing tensors on (A)dS, {\it i.e.}, traceless symmetric tensors that satisfy the Killing equation, \be
\nabla_{(\mu_{t+1}}K^{(0)}_{\mu_{1}\cdots\mu_{t})}=0 \,,
\ee
which implies that they are also divergenceless.  Taken together, these form the  higher-spin algebra of Vasiliev theory \cite{Vasiliev:2003ev,Bekaert:2005vh,Joung:2014qya}, and
from the boundary perspective these parametrize conformal Killing tensors which are the global symmetries of the dual free scalar \cite{Eastwood:2002su}.
For $k\geq1$, these are generalizations of the Killing tensors on (A)dS.  Subsets of them form the higher-spin algebras underlying generalizations of Vasiliev theory containing PM fields \cite{Bekaert:2013zya,Basile:2014wua,Alkalaev:2014nsa,Joung:2015jza,Brust:2016zns}.
Going to the boundary, these parametrize higher-order conformal Killing tensors of the boundary which are global symmetries of higher-derivative dual free scalar theories \cite{2006math.....10610E,2009arXiv0911.5265G,Brust:2016gjy}.

\subsection{Branching rules}
\begin{figure}
\begin{center}
\epsfig{file=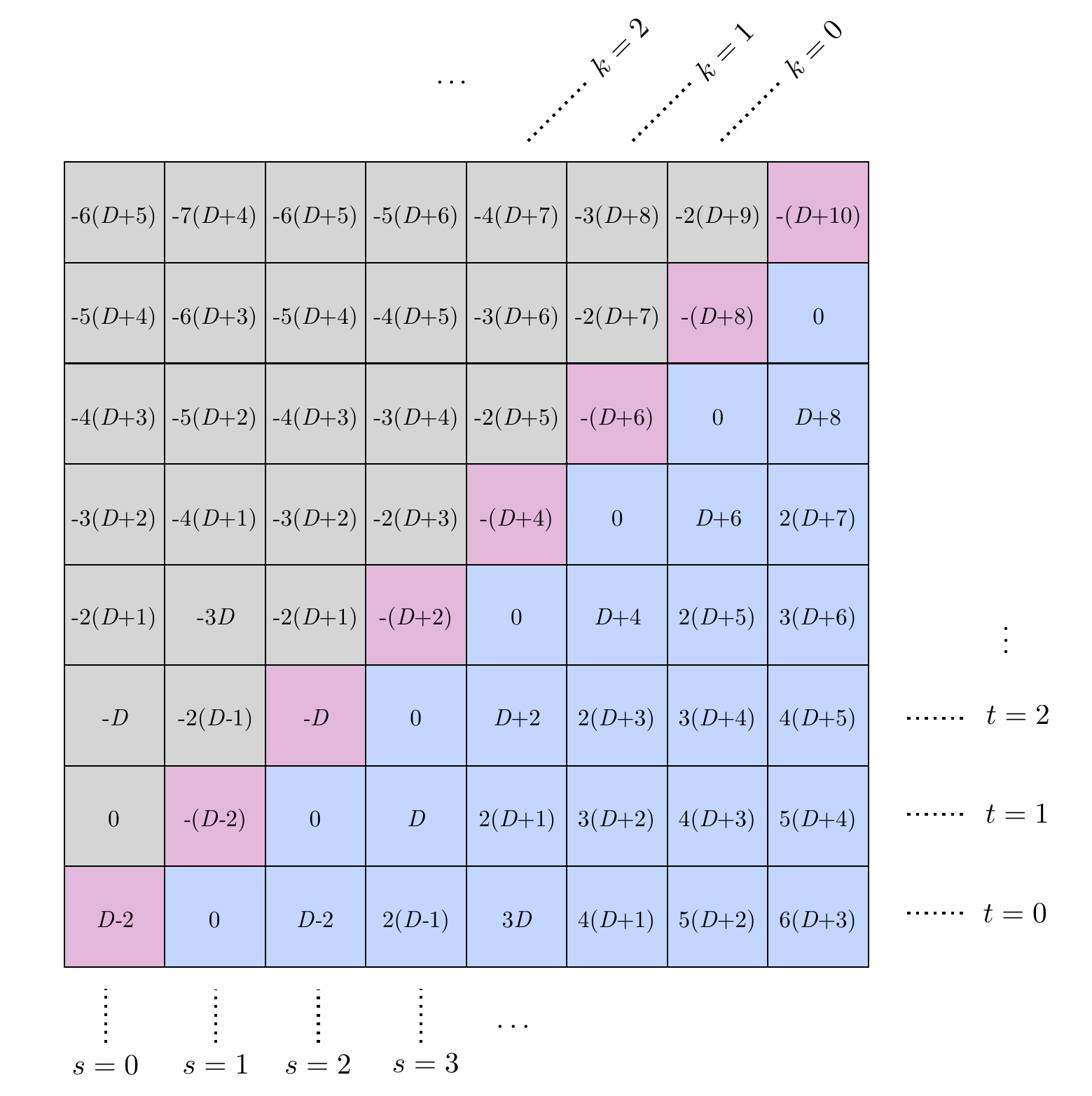,width=4.5in}
\caption{\small Masses in dS space in units of $H^2$ (masses in AdS space are the negatives of these, in units of $1/L^2$).  Theories along the (red) diagonal have no apparent symmetry in general, but the $s=0$ case in $D=4$ corresponds to the conformally coupled scalar.  Each (blue) square below the diagonal is a PM field labelled by spin $s$ and depth $t$.  The corresponding longitudinal mode is the theory reflected across the diagonal, and has a shift symmetry of level $k$. The parameter $k$ is uniform along diagonal lines in the top half of the square.}
\label{masstablegeneralD}
\end{center}
\end{figure}

The shift-symmetric fields introduced in Section~\ref{linesymmsectn} can be understood as the decoupled longitudinal modes of a massive field in the limit that its mass approaches a PM value.
This is analogous to how lower-helicity degrees of freedom are isolated in the massless decoupling limits of theories of massive vector fields or massive gravity~\cite{ArkaniHamed:2002sp}.
When the mass of a massive spin-$s$ field approaches the depth-$t$ PM value \eqref{pmpoints}, its degrees of freedom split into the PM field and an additional massive field that we call the longitudinal mode,
\be (m^2,s)  \xrightarrow[m^2\rightarrow \bar m^2_{s,t}]{} (\bar m^2_{s,t},s)\oplus (m^2_{t,k},t)\, ,\label{genbranchrulemasse}\ee
where $ k=s-1-t$ and the masses of the additional fields are given by
\be
m^2_{t,k}=\begin{cases}-k ( k +D-1)H^2   , & t=0, \\ -(k+2) (k+D-3+2 t)H^2, & t\geq 1\,. \end{cases}
\ee
These longitudinal modes have precisely the masses \eqref{scalarmassvaluese} and \eqref{spinsmassvalueseqe} at which a spin-$t$ field acquires a level-$k$ shift symmetry.  These masses are illustrated in Figure \ref{masstablegeneralD}.

\begin{figure}
\begin{center}
\epsfig{file=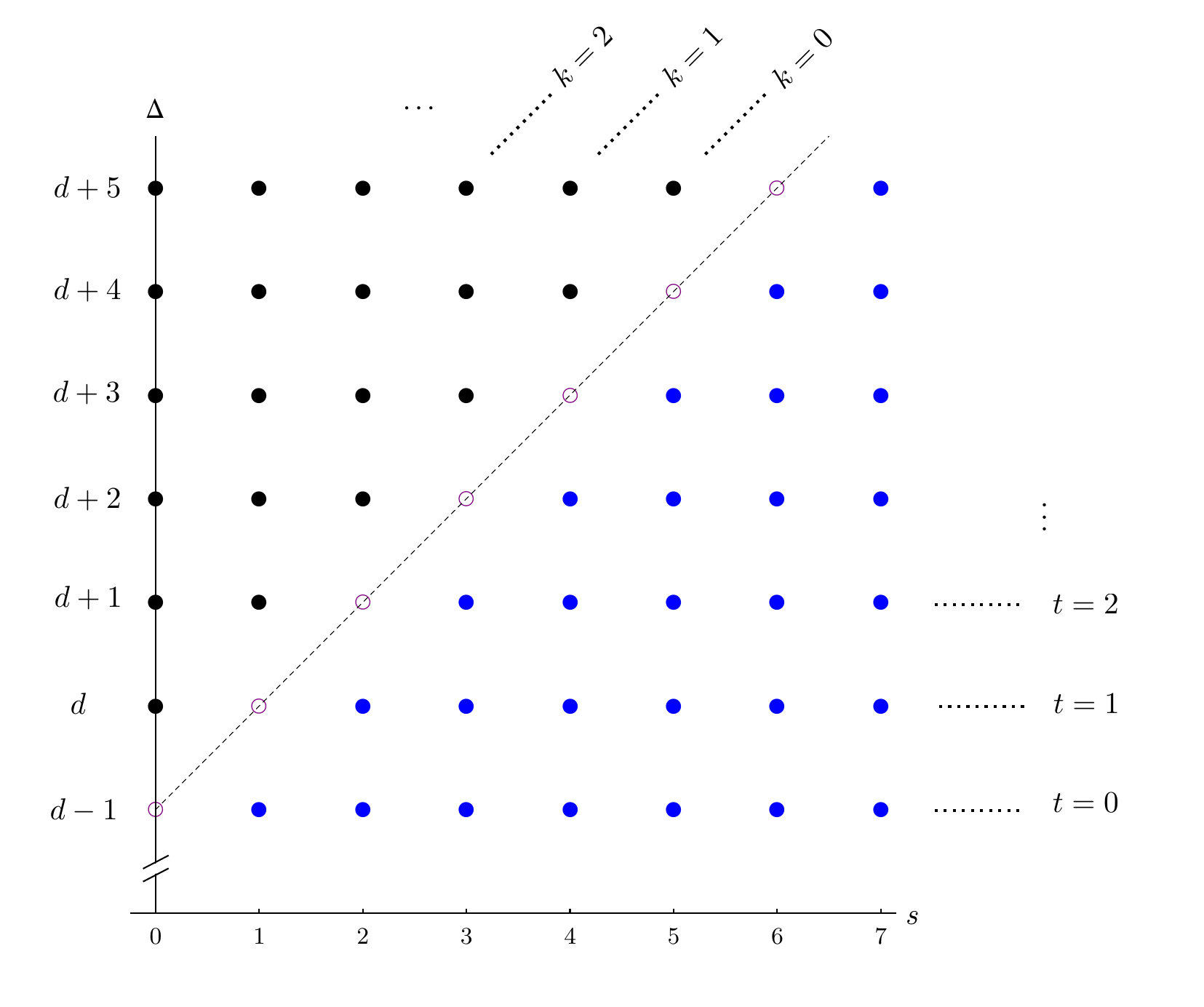,width=5.0in}
\caption{\small Conformal dimensions of the PM and shift-symmetric fields.  Each (blue) filled circle below the diagonal line is a PM field and the corresponding shift-symmetric longitudinal mode is the (black) circle obtained by reflecting about the dotted line $\Delta=s+d-1$.  The open circles on the dotted line belong to the same family as the conformally coupled scalar in $D=4$, corresponding to the diagonal squares in Figure \ref{masstablegeneralD}.}
\label{conformaldimensions}
\end{center}
\end{figure}

This branching rule can be understood in a straightforward way by 
appealing to the representations of the dual CFT in $d=D-1$ dimensions.\footnote{Though we phrase this in terms of AdS representations, identical considerations apply in the dS case.}  We label the CFT representations as $(\Delta,s)$, where $\Delta$ is the scaling dimension of the dual CFT operator given in terms of the mass by the greater root of \eqref{AdSCFTformpe}.  The branching rule then reads
\be \label{eq:cftbranching}
 (\Delta,s) \xrightarrow[\Delta\rightarrow t+d-1]{}  (t+d-1,s)\oplus (s+d-1,t).
\ee
The longitudinal module $(s+d-1,t)$ is precisely the submodule descended from the state \eqref{bosonmultiplyconsce}---obtained by taking $s-t$ divergences---which is going null in the PM limit.  Note that the state $(s+d-1,t)$ is the reflection of the state $(t+d-1,s)$ about the line $\Delta=s+d-1 $ in the $(s,\Delta)$ plane.  This is illustrated in Figure \ref{conformaldimensions}.

The longitudinal mode is a massive field that inherits a shift symmetry.
This shift symmetry comes from the reducibility parameters of the PM gauge transformation discussed in Section~\ref{sec:killingtensors}.  To illustrate this explicitly, consider the PM limit $m^2\rightarrow 2H^2$ of a massive spin-2 field in $D=4$ dS space (see \cite{deRham:2018svs} for the extension of this limit to interacting massive gravity).  The massive spin-2 field, $h_{\mu\nu}$, initially has no gauge symmetry.  At $m^2=2H^2$, it acquires the $t=0$ PM symmetry, 
\be
\delta h_{\mu\nu}= \left(\nabla_\mu \nabla_\nu +H^2 g_{\mu\nu}\right) \xi,
\ee
which removes its longitudinal scalar degree of freedom.  To take the PM limit smoothly in the action, we must introduce a \stu scalar $\phi$ patterned after the gauge symmetry,
\be
h_{\mu\nu}\mapsto h_{\mu\nu}+{1\over H\epsilon}\left( \nabla_\mu \nabla_\nu+H^2 g_{\mu\nu} \right)\phi\,. \label{PMlinestue}
\ee
Here we have defined the quantity
\be \epsilon^2 \equiv m^2-2H^2\, ,
\ee
so that $\epsilon=0$ corresponds to the PM value.  The normalization in front of $\phi$ in \eqref{PMlinestue} is chosen to ensure that $\phi$ is canonically normalized in the PM limit
\be
\epsilon\rightarrow 0,\qquad \qquad H~~~{\rm fixed},\qquad \qquad h\mn, \  \phi \,~~~{\rm fixed}\, . \label{pmdeclimlineare}
\ee
In this limit, the free massive spin-2 action becomes a decoupled PM graviton, $h_{\mu\nu}$, and a scalar, $\phi$, with mass
\be  m_\phi^2=-4H^2 ,\label{PMlongmasse}\ee
as expected from the general branching rule \eqref{genbranchrulemasse}.  The decoupling limit is smooth since before the limit there are five degrees of freedom in the massive spin-2 field and after the limit there are four degrees of freedom in the PM graviton and one in the massive scalar.   

The scalar $\phi$ with the mass \eqref{PMlongmasse} has the $k=1$ shift symmetry.
To see the origin of this shift symmetry, consider the theory after making the \stu replacement \eqref{PMlinestue}, which before taking the PM limit has a \stu gauge symmetry
\be 
\delta h_{\mu\nu}= \left(\nabla_\mu\nabla_\nu+ H^2g_{\mu\nu}\right)\chi,\qquad\qquad\delta  \phi=-H \epsilon\,\chi. \label{PMstymsrese}
\ee
In the limit \eqref{pmdeclimlineare}, the gauge variation of the scalar goes to zero and \eqref{PMstymsrese} becomes the PM gauge symmetry of the PM graviton.  For a generic gauge transformation, we are not free to rescale the gauge parameter so that a nontrivial $\delta\phi$ survives the PM limit, since otherwise the gauge variation $\delta h_{\mu\nu}$ would diverge in this limit. 
There is, however, one exception to this.  If $\chi$ is a reducibility parameter,
\be
\chi=K^{(1)}, \quad \text{with} \quad \left(\nabla_\mu\nabla_\nu +{H^2} g_{\mu\nu}\right)K^{(1)}=0, 
\ee
then the gauge variation $\delta h_{\mu\nu}$ vanishes. We are then free to rescale $K^{(1)}\rightarrow -{1\over H\epsilon}K^{(1)}$ so that $\delta\phi$ survives the PM limit and \eqref{PMstymsrese} becomes 
\be 
\delta h_{\mu\nu}=0,\qquad\qquad \delta  \phi=K^{(1)} = S_AX^A\big|_{\rho =\mathcal{R}}\, , \label{PMstymsrese2}
\ee
which is precisely the $k=1$ shift symmetry of a scalar.  In this way, the reducibility parameters become global symmetries of the \stu field Lagrangians.
The same argument holds for all spins and PM limits.  In Appendix \ref{spin3Appendix}, we illustrate this for a massive spin-3 field, showing explicitly that fields with the expected masses appear in the PM decoupling limits.

\subsection{Unitarity and the Higuchi bound}
\label{sec:unitarity}

The relation between the shift-symmetric fields and PM fields gives us some physical insight into the unitarity of these representations on dS space. Thought of as AdS representations, we can check directly from~\eqref{unitboundd} that all of these fields lie above the unitarity bound. In the dS case, the question of unitarity is more subtle, but we can consider the PM decoupling limit discussed in this section. For massive fields on dS space with $s\geq 2$, there is a lower bound on the mass in order for the field to be unitary---the so-called Higuchi bound---which coincides with the $t=0$ PM point:
\be
m^2 \geq (s-1)(s+D-4)H^2,~~~~~~~~~~{\rm Higuchi~bound}.
\ee
Below this mass value, various helicity components become ghosts. The PM fields with $t>0$ are below the Higuchi bound but are unitary because these ghostly components are projected out. 

We can now imagine taking the PM decoupling limit starting with a higher-spin field with a mass in the healthy region above the Higuchi bound. As we approach the $t=0$ PM point, the representation splits into a depth-$0$ PM field and a scalar field with a shift symmetry of level $k=s-1$. Since we started in the unitary region and the PM field itself is unitary, we expect that the tachyonic scalar is also a unitary representation, by continuity.\footnote{Note that if we take the $t=0$ decoupling limit from below the Higuchi bound, the theory we start with is non-unitary, but this manifests as an overall sign flip of the scalar mode in the decoupling limit, which is consistent with the scalars being unitary.}

In order to reach the $t>0$ PM points, corresponding to shift-symmetric fields with nonzero spin, a similar limiting procedure must first pass {\it below} the Higuchi bound, so the massive representation we start with is not unitary. Since the PM points themselves are unitary, in this case continuity suggests that the decoupled tachyonic fields are not unitary on dS space. In this case, the non-unitarity is a result of relative signs between the helicity components of the shift-symmetric field.

To summarize, all of the shift-symmetric fields discussed in Section~\ref{linesymmsectn} are unitary in AdS space, while we expect that only the scalar fields are unitary in dS space.

\section{Interacting scalars}
\label{sec:interactingscalars}

So far we have only considered shift symmetries of free theories. In this section, we move on to studying interactions that preserve the shift symmetries. 
We will consider interactions of a single scalar field, leaving the interactions of fields with nonzero spin and multiple fields to future work.   
We first look for possible deformations of the symmetry algebras that might be realized by scalar field theories. We then look for theories realizing either the undeformed or deformed symmetry algebras. For $k=1$ and $k=2$, we find interesting interacting examples, including the (A)dS analogue of the special galileon. 

\subsection{Deformed symmetry algebras\label{algebrassection}}

We start by reconsidering the symmetry algebras formed by the generators of shift symmetries and (A)dS isometries.  Interactions can be classified according to whether or not they deform the symmetry algebra of the free theory. In the undeformed case, we call the theories, in a slight abuse of the term, {\it abelian}, otherwise we call them {\it non-abelian}. 

The (A)dS isometries always takes the same ambient space form, \eqref{eq:adssymmtrans}, regardless of interactions. Specialized to scalars, the transformation is
\be 
\delta_{J_{AB}}\Phi \equiv J_{AB}\Phi=X_A\partial_B\Phi-X_B\partial_A\Phi \,.
\ee
It can be checked that these satisfy the $\frak{so}(D+1)$ commutation relations,
\be \left[ J_{AB},J_{CD}\right]= \eta_{AC}J_{BD}-\eta_{BC}J_{AD}+\eta_{BD}J_{AC}-\eta_{AD}J_{BC} \,.
\label{eq:Jcomm}
\ee
In contrast, the traceless shift symmetries acting on the scalar may acquire additional field-dependent terms in the presence of interactions, schematically of the form
\be 
\delta_{S_{A_1\cdots A_k}}\Phi \equiv S_{A_1\cdots A_k}\Phi=X_{(A_1}\cdots X_{A_k)_T}+{\cal O}\left(\Phi\right),
\ee
where the additional terms are constrained to have the same homogeneity degree as $\Phi$.
These additional terms can deform the abelian algebra of the free theory into a non-abelian algebra.
The commutators between the $J_{AB}$ and the $S_{A_1\cdots A_k}$ are fixed because $S_{A_1\cdots A_k}$ is an $\mathfrak{so}(D+1)$ tensor,
\be \label{eq:SJcomm}
\left[J_{BC},S_{A_1\cdots A_{k}}\right] =  \sum_{i=1}^{k} \left( \eta_{BA_i}S_{A_1 \dots A_{i-1} C A_{i+1}\dots A_{k}} - \eta_{CA_i}S_{A_1 \dots A_{i-1} B A_{i+1}\dots A_{k}} \right)\,.
\ee
This leaves only the commutators between  $S_{A_1\cdots A_k}$ to be determined. 
There is only one term consistent with both the symmetries and trace conditions of this commutator, assuming there are no additional generators,\footnote{Naively, when $k$ is even there could also be terms on the right-hand side of this commutator proportional to $S_{A_1\ldots A_i B_{i+1} \ldots B_k}$, but these turn out to be incompatible with the antisymmetry and trace conditions. Such terms are allowed for general higher-spin algebras.}
\be \label{eq:SScomm}
[ S_{A_1 \ldots A_k}, S^{B_1 \ldots B_k} ] =  \alpha k!^2\sum_{n=0}^{\left \lfloor{\frac{k-1}{2}}\right \rfloor} a_n \eta_{(A_1 A_2} \eta^{(B_1 B_2} \ldots \eta_{A_{2n-1} A_{2n}} \eta^{B_{2n-1} B_{2n}} \delta_{A_{2n+1}}^{\ B_{2n+1}} \ldots \delta_{A_{k-1}}^{\ B_{k-1}}J_{A_k)}^{ \ \ B_k)} .
\ee
The $a_n$ are functions of $D$ and $k$ that are fixed by the tracelessness conditions and can be determined recursively by
\be
a_{n+1}= -a_{n} \frac{(k-2n-1)(k-2n-2)}{2(D+2k-2n-4)(n+1)} \,,
\ee 
with $a_0=1$. 

This alone is not enough to guarantee the algebra exists, since we also have to check that the Jacobi identities are satisfied. A nontrivial identity is 
\be \label{eq:jacobi}
\left[S_{A(k)}, \left[S_{B(k)}, S_{C(k)}\right]\right]+\left[S_{B(k)}, \left[S_{C(k)}, S_{A(k)}\right]\right]+\left[S_{C(k)}, \left[S_{A(k)}, S_{B(k)}\right]\right]=0 \,,
\ee
where we have used the condensed index notation, $A(k) \equiv A_1\cdots A_k$.
Using \eqref{eq:SScomm} and \eqref{eq:SJcomm}, the first term is given schematically by
\be \label{eq:jacobi1}
[S_{A(k)}, [S_{B(k)}, S_{C(k)}]] \sim \alpha \! \sum_{n=0}^{\left \lfloor{\frac{k-1}{2}}\right \rfloor} a_n (\eta_{BB})^n  (\eta_{CC})^n  (\eta_{BC})^{k-2n-1} \!\left(\eta_{CA} S_{B A(k-1)}-\eta_{BA} S_{C A(k-1)} \right),
\ee
where each index type is to be separately symmetrized. Permuting the indices gives the other two terms in \eqref{eq:jacobi}.
For $k>2$, the sum in \eqref{eq:jacobi1} contains terms with $n>0$ that, due to their index structure, cannot cancel against other terms in \eqref{eq:jacobi}, so the Jacobi identities can only be satisfied if $\alpha =0$. This implies that the $k>2$ abelian algebras cannot be deformed without introducing additional particles, at least in generic dimensions. There could still be algebras with $k>2$ that exist only in specific dimensions due to dimension-dependent identities, {\it e.g.}, as in \cite{Manvelyan:2013oua}. Additionally, in $D\leq 3$ there is the possibility of parity-violating terms in the algebras, but we do not consider such cases.

For $k\leq 2$, we can check explicitly that all of the Jacobi identities are satisfied for any value of $\alpha$, so deformations of the algebra can exist. We next discuss each of these cases.

\subsubsection{$k=0$ algebra}

When $k=0$, the generator $S$ is a scalar and hence always commutes with itself, just as in the free theory.  The symmetry algebra is simply $\frak{so}(D+1)\oplus \frak{u}(1)$ and there are no non-abelian extensions without introducing additional generators.
It is straightforward to write down theories in (A)dS space realizing this algebra. Any interactions where at least one derivative appears on each field will be invariant,  which includes ghost-free theories such as $P(X)$ theories. The tadpole term is also invariant and is a Wess--Zumino term for this shift symmetry.

\subsubsection{$k=1$ algebra}

When $k=1$, the commutator \eqref{eq:SScomm} is given by
\be 
\left[S_A,S_B\right]=\alpha J_{AB} \, .\label{k1commutatoralge}
\ee
In the abelian theory, $\alpha=0,$ the algebra is $\frak{iso}(D+1)$, with $S_A$ playing the role of the translations.  
We can therefore think of the abelian $k=1$ scalar as the Goldstone field for the symmetry breaking pattern
\be \label{eq:k1linearSB}
\frak{iso}(D+1)\longrightarrow \frak{so}(D+1) \,.
\ee
In Section~\ref{sec:k1int} we describe nonlinear interactions which realize this symmetry algebra.

For $\alpha\not=0$, the algebra is $\frak{so}(D+2)$.  This can be seen by grouping $J_{AB}$ and $S_A$ into a $(D+2)$-dimensional antisymmetric matrix with $S_A$ along the first row and column.
We can therefore think of scalar fields invariant under this deformed algebra as realizing the symmetry breaking
\be \label{eq:conformalbreaking}
\frak{so}(D+2)\longrightarrow \frak{so}(D+1) \,.
\ee
The algebra \eqref{k1commutatoralge} can be realized on a spin-0 field by the ambient space transformation
\be
\delta_{S_A} \Phi \equiv S_A\Phi=X_A+\alpha\Phi \partial_A \Phi.\label{k1nonlinetranse}
\ee
In Section~\ref{sec:k1int} we construct interactions with second-order equations of motion that are invariant under this transformation and discuss other ways of realizing the symmetry breaking \eqref{eq:conformalbreaking}.

\subsubsection{$k=2$ algebra}
When $k=2$, the commutator \eqref{eq:SScomm} is given by
\be
[S_{A_1 A_2}, S_{B_1 B_2}] = \alpha \left( \eta_{A_1 B_1} J_{A_2 B_2} + \eta_{A_2 B_1} J_{A_1 B_2}+ \eta_{A_1 B_2} J_{A_2 B_1} + \eta_{A_2 B_2} J_{A_1 B_1} \right).
\label{k2commutatoralge}
\ee
When $\alpha \not=0$, we can combine the generators into a matrix
\be 
M_{AB}\equiv -\frac{1}{2}J_{AB}\pm \frac{i}{2 \sqrt{\alpha}}S_{AB},
\ee
where the tracelessness of $S_{AB}$ implies that $\eta^{AB} M_{AB}=0$.
The commutators written in terms of $M_{AB}$ are
\be \left[M_{AB},M_{CD}\right]= \eta_{BC} M_{AD} - \eta_{AD} M_{CB},\ee
which are the commutators of $\frak{sl}(D+1)$.  
In a scalar theory realizing this symmetry algebra, the $J_{AB}$ are linearly realized and the shift symmetries $S_{AB}$ are non-linearly realized, so we may think of the scalar field as the Goldstone field for the symmetry breaking
\be
\frak{sl}(D+1)\longrightarrow\, \frak{so}(D+1).
\ee
This nonlinear symmetry can be realized on a spin-0 field by the ambient space transformation
\be
\delta_{S_{AB}}\Phi\equiv S_{AB}\Phi = X_{(A} X_{B)_T}+\alpha \partial_{(A}\Phi\partial_{B)_T}\Phi, \label{nonlinek2syme}
\ee
which is just the  flat space special galileon transformation~\cite{Hinterbichler:2015pqa}.  In Section \ref{specialgalileonadssec}, we present the unique ghost-free (A)dS theory invariant under the transformation \eqref{nonlinek2syme}, which is the curved space analogue of the special galileon.

\subsection{$k=1$ interactions}
\label{sec:k1int}
Now that we have the possible symmetry algebras that a $k=1$ scalar can realize, we look for invariant interactions. In this case, the theories have been considered before in the literature in slightly different contexts.
These scalar fields can be thought of as Goldstone degrees of freedom associated with the nonlinearly realized symmetries, and can be constructed in a systematic way using nonlinear realization techniques.  In Appendix~\ref{cosetappendix}, we employ this method to construct interactions for the abelian $k=1$ and $k=2$ scalars.

\subsubsection{Abelian interactions}
\label{sec:abeliank1}
We first look for a scalar theory that realizes the $k=1$ abelian symmetry algebra. In this case, the shift symmetries commute and act on the ambient space field as 
\be
\delta\Phi = S_A X^A.
\ee
By acting on the scalar with two ambient space derivatives and projecting using the rules outlined in Appendix~\ref{embbedappendix}, we get the invariant ``curvature" tensor
\be
\partial_A\partial_B\Phi \leadsto \Phi^{(1)}_{\mu\nu} = \left(\nabla_{\mu } \nabla_{\nu} + H^2 g_{\mu \nu}\right) \phi.
\label{eq:k1invariant}
\ee
We can construct invariant interactions by forming scalars from this tensor.

Since the tensor $\Phi^{(1)}_{\mu\nu}$ has a term with two derivatives, most interactions lead to higher-derivative equations of motion and ghosts.  There are, however, a set of $D+1$ terms that are ghost free.  These are the (A)dS galileons, which were discovered in~\cite{Goon:2011uw,Goon:2011qf,Burrage:2011bt} and appear in the PM decoupling limit of massive gravity~\cite{deRham:2018svs}.  They are listed in Eqs.~\eqref{eq:4AdSGals} and~\eqref{eq:5AdSGal}. In Appendix \ref{cosetappendix}, we perform a coset construction of the (A)dS galileons, showing that $D$ of them are constructible from the invariant tensor \eqref{eq:k1invariant}. The last interaction is a Wess--Zumino term, which cannot be written solely in terms of $\Phi^{(1)}_{\mu\nu}$.  Another realization of the symmetry breaking pattern \eqref{eq:k1linearSB} comes from considering a dS brane in a flat bulk~\cite{Goon:2011uw,Goon:2011qf}. This theory should be related to the (A)dS galileons by a (possibly nonlocal) field redefinition.\footnote{This is because both theories realize the same symmetry breaking pattern. It is expected that theories of Goldstone bosons are essentially unique and that all nonlinear realizations of the same symmetries with the same degrees of freedom are equivalent. This has been proven in the case of internal symmetries~\cite{Coleman:1969sm,Callan:1969sn}, but remains conjectural in the case of spacetime symmetries. 
}

\subsubsection{Non-abelian interactions}
We now consider interactions that are invariant under the non-abelian $k=1$ algebra, $\mathfrak{so}(D+2)$. This algebra can be realized on a scalar field through the ambient space transformation
\be \label{eq:k=1transform}
\delta \Phi = S_A X^A + \frac{1}{\Lambda^D}S_A \Phi \partial^A \Phi.
\ee
This is the same as \eqref{k1nonlinetranse} but with $\alpha$ now written in terms of the dimensionful scale $\Lambda$, which is not to be confused with the cosmological constant. In $D$ dimensions there are $D+1$ ghost-free interactions invariant under \eqref{eq:k=1transform}, the $n$\textsuperscript{th} of which can be written as\footnote{To find these interactions, we started with a general ansatz with second-order equations of motion, constructed with a judicious choice of variables. We then imposed order by order the subset of symmetries \eqref{eq:k=1transform} for a field depending only on the Poincar\'e coordinates $(\eta,x^1)$, up to a very high order, and then we resummed the result.}
\begin{align}
\frac{\mathcal{L}_n }{\sqrt{-g}}&= \sum_{j=0}^{D-1}\sum_{m=1}^{D-j}   \frac{c_{j, m,n}\bar{\phi}^{m-1}}{(\bar{\phi}^2-1)^{D/2+1}} \left[ (j+2) \tilde{f}_{j,m}\left(\frac{X}{\bar{\phi} ^2-1} \right)-(j+1) \tilde{f}_{j+1,m-1}\left(\frac{X}{\bar{\phi} ^2-1} \right) \right]  \partial^{\mu} {\phi} \partial^{\nu} {\phi} X^{(j)}_{\mu \nu}(\Pi) \nonumber \\
& + \Lambda^D V_n( \bar{\phi}), \label{eq:k1L}
\end{align}
where
\be
\bar{\phi} \equiv -\frac{iH\phi}{\Lambda^{D/2}}, 
\quad X \equiv \frac{\partial_{\mu} \phi \partial^{\mu} \phi }{ \Lambda^{D}},
\ee
and
\be
\tilde{f}_{j,m}(x) \equiv  \, _2F_1\left(\frac{j+1}{2},\frac{j+m+2}{2} ;\frac{j+3}{2}; x\right),
\ee
\be
 c_{j, m,n} \equiv \frac{ i^{D+m+1} (j+m-1) \left((-1)^{m+n+j}-1\right) \Gamma \left(\frac{D+4}{2}\right) \Gamma \left(\frac{j+m+2}{2}\right) (2-j-m)_{j-1} }{H^j\Lambda^{D j/2} D(D-1) (j+2)! \, \Gamma
   \left(\frac{j+m}{2}\right) \Gamma \left(\frac{ j+m-n+3}{2}\right) \Gamma \left(\frac{D-j-m+n+3}{2}\right)(2-D)_{j-1}}.
\ee
The tensors $X^{(j)}_{\mu\nu}$ are defined in Appendix \ref{xtensorappendix} and depend on the matrix $\Pi_{\mu\nu}\equiv\nabla_\mu\nabla_\nu\phi$.
The potentials can be written as
\be
V_n( \bar{\phi}) = -\int \rd \bar{\phi} \sum_{m=1}^{D+1} c_{-1,m,n} \frac{\bar{\phi}^{m-1} }{(\bar{\phi}^2-1)^{D/2+1}}.
\ee
The general Lagrangian is
\be
\mathcal{L} =\sum_{n=1}^{D+1} a_n \mathcal{L}_n,
\ee
where $a_n$ are real constants. To canonically normalize the kinetic term we set $a_2=D$ and to remove the tadpole term we set $a_1=0$.

The Lagrangians~\eqref{eq:k1L} are written for a particular choice of field variables in which the symmetry transformation takes a particularly simple form in embedding space. There are, however alternative choices where the final invariant actions take a simpler form, at the price of complicating the ambient space field transformation. 
In particular, the symmetry breaking pattern \eqref{eq:conformalbreaking} has been considered previously in~\cite{Hinterbichler:2012mv,Creminelli:2012qr} for different reasons. The construction in this case is most simply phrased directly in the physical (A)dS space.  The scalar field is taken to transform as
\be \label{eq:k1nonlinear}
\delta\phi = \frac{1}{D}\nabla_\mu\xi^\mu+\xi^\mu\partial_\mu\phi,
\ee
where $\xi^\mu$ are the conformal Killing vectors of (A)dS space.\footnote{The explicit form of these symmetry transformations in the flat slicing can be found in~\cite{Hinterbichler:2012mv}}
Although this transformation realizes the same algebra as~\eqref{eq:k=1transform}, it is not of the form of a deformed ambient space polynomial and the invariant interactions differ from~\eqref{eq:k1L}. 
To find interactions invariant under \eqref{eq:k1nonlinear}, we can define an object $\bar R_{\mu\nu}$ that transforms like a tensor under this transformation,
\be
\bar R_{\mu\nu} = (D-1)H^2 g_{\mu\nu}-(D-2)\nabla_\mu\nabla_\nu\phi-g_{\mu\nu}\Box\phi+(D-2)\partial_\mu\phi\partial_\nu\phi-(D-2) g_{\mu\nu}(\partial\phi)^2,
\ee
where $g_{\mu\nu}$ is the (A)dS metric.
Invariant interactions are then obtained as scalars built out of $\bar R_{\mu\nu}$ and the metric $\bar g_{\mu\nu}  = e^{2\phi}  g_{\mu\nu}$, for which $\bar R_{\mu\nu}$ is the Ricci curvature. For example, one invariant interaction is given by the Lagrangian
\be
{\cal L} = \sqrt{-\bar g}\left(-\bar R+(D-1)(D-2)H^2\right),
\ee
which results in the expected mass $m^2 = -DH^2$. There is also a Wess--Zumino term that cannot be constructed from these covariant building blocks and whose explicit form is given in~\cite{Hinterbichler:2012mv}, along with the details of the coset construction. 
There is also a DBI presentation of this symmetry breaking pattern~\cite{Goon:2011uw,Goon:2011qf}. It seems likely that there are field transformations that link these different forms, as in flat space~\cite{Bellucci:2002ji, Creminelli:2013ygt}.

\subsection{$k=2$ interactions}
We now consider interactions for a $k=2$ scalar. Similar to $k=1$, there exist abelian interactions built from a shift-invariant curvature. We also find a theory that realizes the non-abelian extension of the $k=2$ symmetry algebra, which is a curved space version of the special galileon.

\subsubsection{Abelian interactions}
We first consider theories that nonlinearly realize the abelian symmetry algebra, where the $S_{AB}$ generators commute. 
The symmetries act on the ambient space scalar as
\be
\delta\Phi = S_{AB}X^AX^B.
\label{eq:abeliank2shifts}
\ee
Using ambient space, we can construct the following invariant tensors:
\begin{align}
\partial_A\partial_B\partial_C\Phi & \leadsto \Phi^{(2)}_{\mu\nu\rho} = \left(\nabla_{(\mu}\nabla_\nu\nabla_{\rho)}+4H^2 g_{(\mu\nu}\nabla_{\rho)}\right)\phi,
\label{eq:phi2tensor} \\
\partial_A\partial^A \Phi & \leadsto \Phi^{(2)} = \left(\square+2(D+1)H^2\right)\phi.
\end{align}
These tensors are invariant under the restriction of the shifts~\eqref{eq:abeliank2shifts} to (A)dS space, so any scalar formed from them and their derivatives will provide an invariant Lagrangian. There is also the possibility of Wess--Zumino terms which are not directly constructible from these objects. However, as in flat space~\cite{Hinterbichler:2014cwa}, all interactions invariant under the abelian $k=2$ shift symmetry have higher-order equations of motion. Such interactions may still be of interest, for example, to describe the helicity-0 mode interactions of a massive spin-3 particle in (A)dS space. 

\subsubsection{Non-abelian interactions \label{specialgalileonadssec}}

We now consider interactions that are invariant under the non-abelian $k=2$ algebra, $\mathfrak{sl}(D+1)$, where the scalar transformation is given in ambient space by
\be \delta \Phi=S_{AB}\left(X^A X^B +{1\over \Lambda^{D+2} }\partial^A\Phi \partial^B\Phi\right), \label{nonlinek2syme2}
\ee
with symmetric traceless $S_{AB}$.
Using the same method as for the non-abelian $k=1$ interactions, we find that there is a single invariant interaction with second-order equations of motion,\footnote{We have not been able to verify symbolically that~\eqref{fulllagk2nonline} has the full symmetry to all orders, although we are confident it does.  For $D=4$, we have shown that it has the full symmetry up to 14$^{\rm th}$ order in the fields and has the relevant subset of symmetries to all orders when $\phi = \phi(\eta,x^1)$. We have also shown that the expression for the symmetry variation of the full Lagrangian vanishes when random integers are substituted for the variables, and so we are confident it is zero, but due to its size we have not managed to simplify it to zero symbolically.}
\begin{align} \frac{ {\cal L}_{\rm SG}}{ \sqrt{-g}}=& 
\sum_{j=0}^{D-1}{ \psi^{D-j}+(-1)^j{\psi^\ast}^{D-j}\over i^j  \Lambda^{j\left(D +2\right)/2}\left|\psi\right|^{D+3}2\,\Gamma(j+3)} \left[(j+1)f_{j+1}\left({X\over \left|\psi\right|^2}\right)-(j+2)f_j\left({X\over \left|\psi\right|^2}\right) \right]\partial^\mu\phi\partial^\nu\phi X^{(j)}_{\mu\nu}(\Pi) \nn\\
& + \frac{\Lambda^{D+2}}{ 2 (D+1) H^2}\left(1-{{\psi^\ast}^{D+1}+\psi^{D+1}\over 2 \left|\psi\right|^{D+1}}\right), \label{fulllagk2nonline}
\end{align}
where we have defined\footnote{We use the definition of ${}_2F_1$ of \href{http://functions.wolfram.com/HypergeometricFunctions/Hypergeometric2F1}{\tt functions.wolfram.com/HypergeometricFunctions/Hypergeometric2F1}.}
\be \label{eq:hypergeo}
f_j(x) \equiv  {}_2F_1\left({D+3\over 2},{j+1\over 2};{j+3\over 2};-x\right), \qquad \psi\equiv 1-2i {H^2\over \Lambda^{{D\over 2}+1}}\phi\, , \qquad X\equiv {H^2\over \Lambda^{D+2}}(\partial\phi)^2\,.
\ee
Despite the appearance of factors of $i$, this Lagrangian is real. Note that in this case we have ignored the tadpole term, which  can be obtained from \eqref{fulllagk2nonline} by putting minus signs in front of $\psi^{* D-j}$ and $\psi^{D+1}$. There will also exist invariant interactions with higher-order equations of motion, which are (A)dS versions of the interactions constructed in~\cite{Novotny:2016jkh}.

The Lagrangian \eqref{fulllagk2nonline} depends on two scales, $H$ and $\Lambda$, and therefore has only one dimensionless parameter, the ratio of these scales.  Only terms of even order in $\phi$ are present, so there is also a ${\mathbb Z}_2$ symmetry $\phi\rightarrow -\phi$.

In the flat space limit, $H\rightarrow 0$, \eqref{fulllagk2nonline} reduces to
\be {\cal L}_{\rm SG}\Big\rvert_{H=0}=-\sum_{\substack{j=0, \\ j\,{\rm even}}}^{D-1}{1\over \Lambda^{j\left(D+2\right)/2}}{(-1)^{j/2}\over (j+2)!}\partial^\mu\phi\partial^\nu\phi X^{(j)}_{\mu\nu}(\Pi). \ee
Up to a total derivative, this is precisely the Lagrangian for the flat space special galileon of \cite{Cheung:2014dqa,Hinterbichler:2015pqa,Cheung:2016drk}, 
which is invariant under the nonlinear symmetry
\be \delta \phi =s_{\mu\nu}\left(x^\mu x^\nu +{1\over \Lambda^{D+2} }\partial^\mu\phi\partial^\nu\phi\right),
\ee
with symmetric traceless $s_{\mu\nu}$~\cite{Hinterbichler:2015pqa}, as well as the standard galileon and shift symmetries.
The Lagrangian \eqref{fulllagk2nonline} can therefore be considered an $H^2$ deformation of the flat space special galileon.

Expanding in powers of the field, the structure of the Lagrangian~\eqref{fulllagk2nonline} is
\be  {1\over \sqrt{-g}}{\cal L}_{\rm SG}=-{1\over 2}(\partial\phi)^2+(D+1)H^2\phi^2+{1\over 24\Lambda^{D+2}}\left[\partial^\mu\phi\partial^\nu\phi X^{(2)}_{\mu\nu}(\Pi)+{\cal O}\left(H^2\right)\right]+{\cal O}\left(\phi^6\right),
\ee
which has the correct mass for a $k=2$ scalar. Each power of $\phi$ comes suppressed with powers of $\Lambda$, with a tail of lower-derivative terms suppressed by $H$.  These tail terms are different from those of the (A)dS galileons and there are interactions at every even order in the field. Thus, unlike the flat space special galileon, which is a particular combination of galileons, the Lagrangian \eqref{fulllagk2nonline} is {\it not} a particular combination of (A)dS galileons. 

In the non-abelian $k=1$ case, we found that there is a formulation of the interactions that is much simpler than the one obtained from the simple ambient space field transformation. Based on this, we expect that there should also exist a choice of field variables where the non-abelian $k=2$ theory takes a much simpler form than~\eqref{fulllagk2nonline}. It would be interesting to find such a presentation of the theory.

In any given dimension, the hypergeometric functions in Eq.~\eqref{eq:hypergeo} can be written in terms of elementary functions: square roots and  (in odd dimensions) $\tanh^{-1}(\sqrt{x})$. For example, in $D=4$ the Lagrangian \eqref{fulllagk2nonline} can be written as
\begin{align}
{1\over \sqrt{-g}}{\cal L}_{\rm SG}=&  -\frac{\Lambda ^6 }{ H^2}\frac{ (y^2-8 y+8) \left(8 X^2-3 y^{3/2} \sqrt{X+y}+12 X y-3 X \sqrt{y} \sqrt{X+y}+3 y^2\right)}{15y^3
   (X+y)^{3/2}} \nn\\ 
&   -\frac{\Lambda ^6 }{ H^2} \left(\frac{ 5 (y-4) y+16}{10y^{5/2}}-{1\over 10}\right)    
+\frac{2 (y-4) \phi }{15 X y^{5/2}} \left(\frac{\sqrt{y} (2 X+3 y)}{(X+y)^{3/2}}-3\right){H^2\over \Lambda ^6}\partial^\mu\phi\partial^\nu\phi X^{(1)}_{\mu\nu}(\Pi) \nn\\
&+\frac{y-2}{30  X^2 y^2} \left(2 \sqrt{y}-\frac{2 X^2+3 X y+2 y^2}{(X+y)^{3/2}}\right){1\over \Lambda ^6}\partial^\mu\phi\partial^\nu\phi X^{(2)}_{\mu\nu}(\Pi) \nn\\
&+\frac{\phi }{45 X^2 y^{3/2}} \left(\frac{\sqrt{y} (3 X+2 y)}{(X+y)^{3/2}}-2\right){H^2\over \Lambda^{12}}\partial^\mu\phi\partial^\nu\phi X^{(3)}_{\mu\nu}(\Pi),
\end{align}
where we have defined
\be y\equiv 1+4{H^4\over \Lambda^6}\phi^2,\ \ \ X\equiv {H^2\over \Lambda^6}(\partial\phi)^2 \,.\ee
Though this Lagrangian may seem somewhat complex, we emphasize that it is entirely fixed by the shift symmetries~\eqref{nonlinek2syme2}.

\subsubsection*{Potential}

\begin{figure}
\begin{center}
\epsfig{file=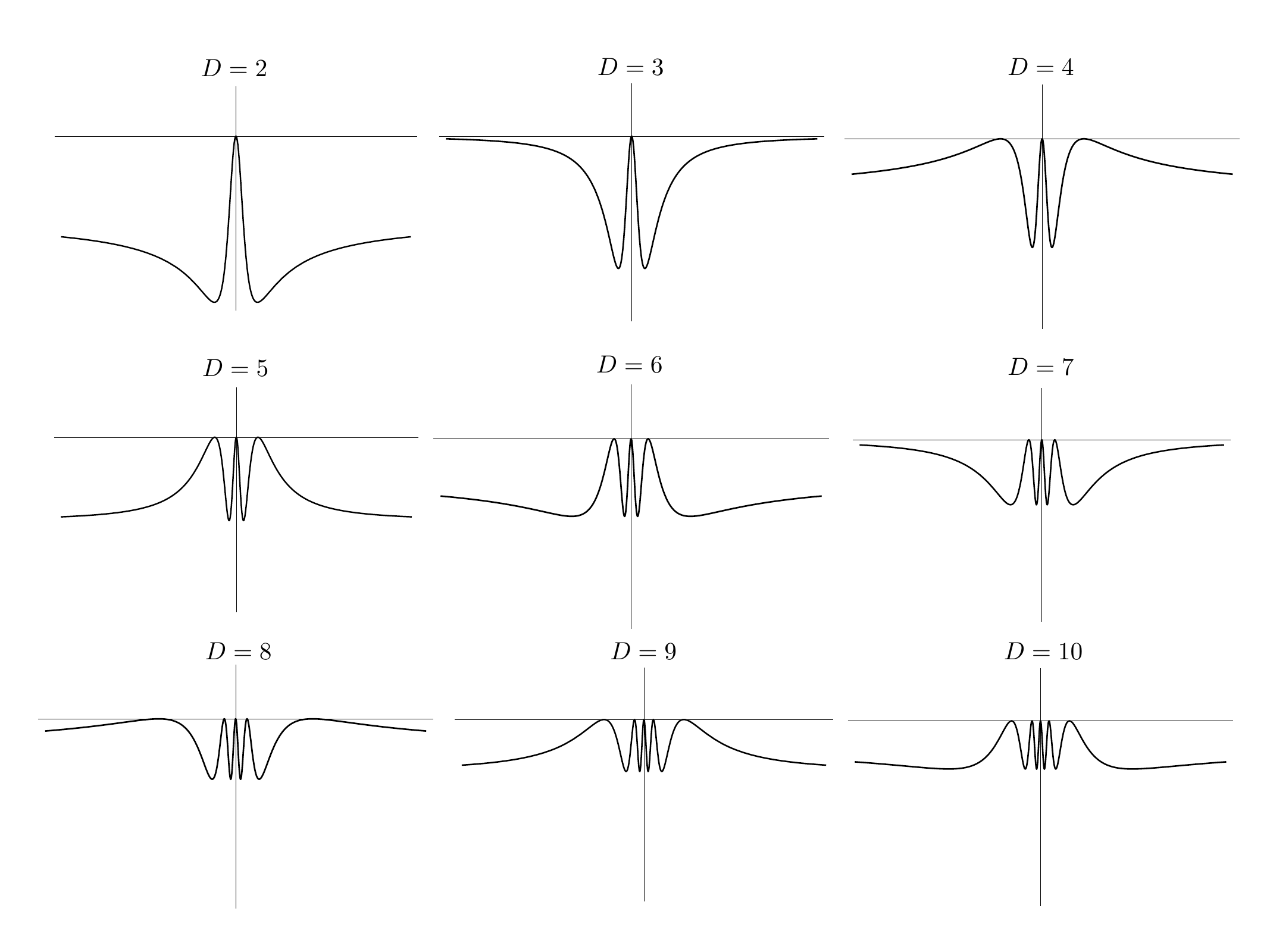,width=6.5in}
\caption{\small Plots of $H^2V(\phi)/\Lambda^{D+2}$ for the special galileon in dS space in various dimensions. The horizontal axes show the dimensionless combination $H^2\phi/ \Lambda^{(D+2)/2}$.}
\label{potentialsfig}
\end{center}
\end{figure}

The (A)dS special galileon possesses a potential whose form is completely fixed by the symmetry. In $D$ dimensions, this potential is given by
\be   V(\phi)=-{1\over \sqrt{-g}}{\cal L}_{\rm SG}\bigg|_{\partial\phi=0}= { \Lambda^{D+2}\over  2(D+1) H^2}\left({\psi^{D+1}+{\psi^\ast}^{D+1}\over 2 \left|\psi\right|^{D+1}} -1\right),
\ee
where we recall that
\be
\psi\equiv 1-2i H^2\phi/ \Lambda^{(D+2)/2}.
\ee
We plot this potential in Figure~\ref{potentialsfig} for $D=2, \ldots, 10$ for the dS case $H^2>0$. The ${\mathbb Z}_2$ symmetry implies that the potential is invariant under interchanging $\psi \leftrightarrow \psi^*$, so the AdS case corresponds to an overall sign flip.
The potential is bounded, with the asymptotic values
\be
V(\phi\to \pm \infty) \simeq - { \Lambda^{D+2}\over H^2}\frac{\sin \left(\frac{\pi  D}{2}\right)+1}{2 (D+1)}.
\ee
The number of critical points is $2\left \lfloor{{D\over 2}}\right \rfloor  +1$, increasing by two every two dimensions, and all the maxima/minima have the same height. 

\begin{figure}
\begin{center}
\epsfig{file=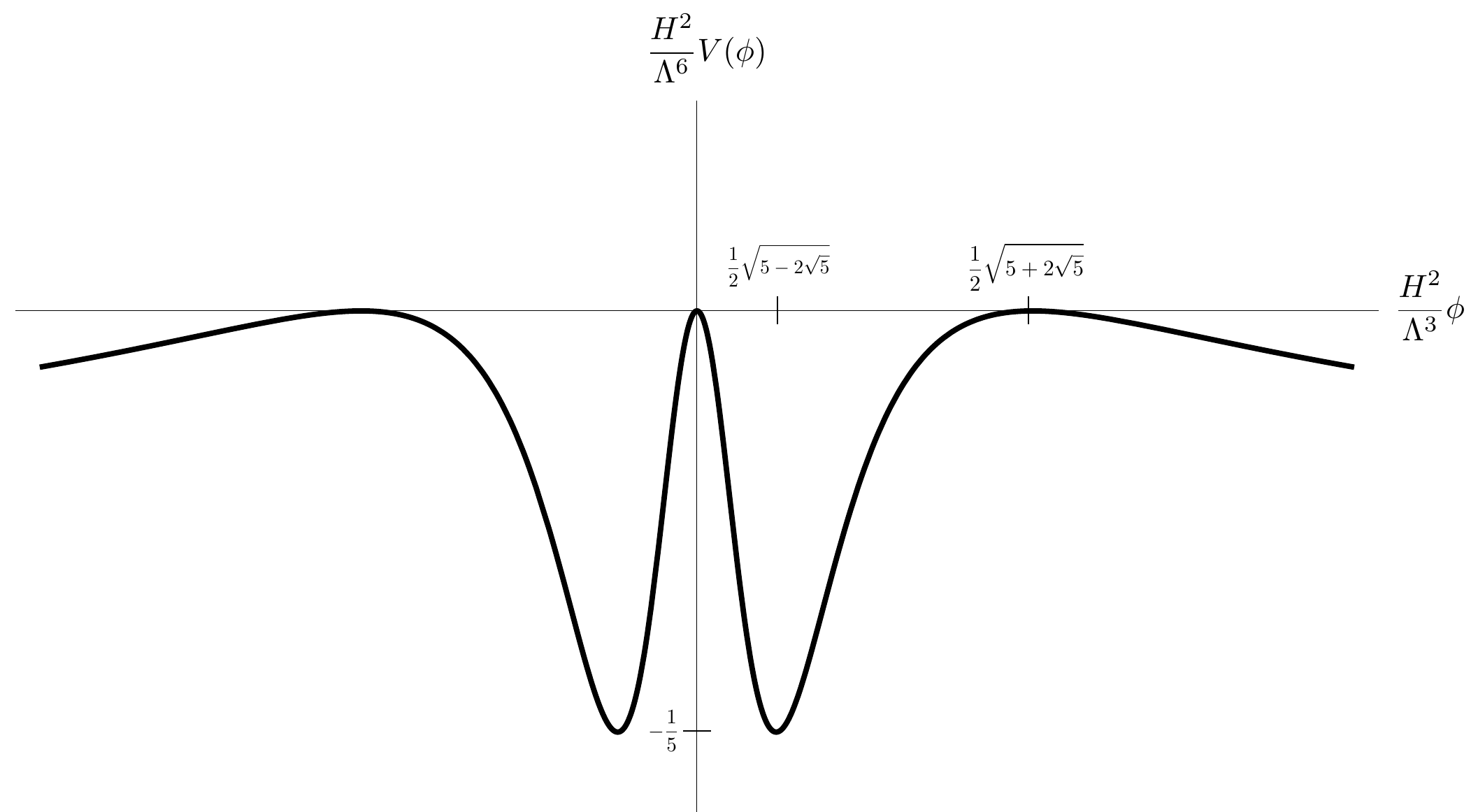,width=4.75in}
\caption{\small Potential for the $D=4$ special galileon in dS space.
}
\label{potential4d}
\end{center}
\end{figure}

Focusing on the case $D=4$, the potential can be written as
\be V(\tilde{\phi})=-{1\over \sqrt{-g}}{\cal L}_{\rm SG}\bigg|_{\partial\phi=0}={\Lambda^6\over 10 H^2}  \left(\frac{80 \tilde{\phi} ^4-40 \tilde{\phi} ^2+1}{ \left(4 \tilde{\phi}^2+1\right)^{5/2}}-1 \right)\,,
   \ee
where $\tilde{\phi} \equiv H^2 \phi/ \Lambda^3$.
This is plotted in Figure~\ref{potential4d}.  
As $\tilde{\phi}\rightarrow \pm \infty$, the potential asymptotes to $-\Lambda^6/10H^2$. There are five critical points, two absolute minima and three absolute maxima,
\begin{align}
\text{Minima: }  \tilde{\phi}_{\rm min}& =\pm  \frac{1}{2} \sqrt{5-2 \sqrt{5}}\ ,\quad  { H^2\over\Lambda^6}V(\tilde{\phi}_{\rm min})=-{1\over 5}\,, \\
\text{Maxima: }\tilde{\phi}_{\rm max} &=0, \ \pm \frac{1}{2} \sqrt{5+2 \sqrt{5}}\ ,\quad V(\tilde{\phi}_{\rm max})=0\, .
\end{align}

Expanding about the critical points, the value of the squared mass relative to the kinetic term is always $-10H^2$, as required by the shift symmetry, which is a nontrivial check of the symmetry of the entire Lagrangian.  However, the overall sign of the quadratic action differs around the maxima and minima,
\begin{align}  {1\over \sqrt{-g}}{\cal L}_{\rm SG} & \propto -\left((\partial\vp)^2-10H^2\vp^2\right)+{\cal O}\left(\vp^3\right),\ \ \ \ \phi=\phi_{\rm max}+\vp \, ,\\
 {1\over \sqrt{-g}}{\cal L}_{\rm SG} & \propto +\left((\partial\vp)^2-10H^2\vp'^2\right)+{\cal O}\left(\vp^3\right),\ \ \ \ \phi=\phi_{\rm min}+\vp.
\end{align}
Although the minima appear stable by considering just the potential, the fluctuations about them have ghostly kinetic terms, so they are actually unstable in dS space. As discussed at the end of Section~\ref{sec:linearscalars},  the would-be growing mode corresponding to the tachyon instability around the origin can be removed by a symmetry transformation. In AdS space, the potential is flipped upside down and there are three stable minima, around which the field has the correct sign kinetic term.


\subsection{$k > 2$ interactions}
We now comment briefly on interactions for shift symmetries with $k>2$. As discussed in Section~\ref{algebrassection}, there do not exist deformations of the symmetry algebras when $k>2$ without adding additional generators, except possibly in special dimensions. This implies that interacting theories of a single scalar in generic dimensions must realize the symmetries of the free theory, which act as
\be 
\delta\Phi=S_{A_1\cdots A_k}X^{A_1}\cdots X^{A_k}.
\label{eq:kg2symms}
\ee
It is straightforward to construct objects that are invariant under these shifts by acting with $k+1$ derivatives or the d'Alembert operator in ambient space and projecting, 
\begin{align}
\partial_{A_1}\dots \partial_{A_{k+1}}\Phi & \leadsto \Phi^{(k)}_{\mu_1\cdots\mu_{k+1}}= \nabla_{(\mu_1}\dots \nabla_{\mu_{k+1})}\phi+{\cal O}(H^2) {\rm \ terms}, \\
\partial_A \partial^A \Phi & \leadsto \Phi^{(k)} =  \left(\square+k(D+k-1)H^2\right)\phi.
\end{align}
Any scalar formed from these objects and their derivatives will be an invariant Lagrangian. This captures all possible Lagrangians except for a finite number of Wess--Zumino terms. 
It is beyond the scope of this work, but it would be interesting to classify such Wess--Zumino terms for these higher shift symmetries. Our expectation is that all interactions invariant under the symmetries~\eqref{eq:kg2symms} with $k\geq2$ will have higher-order equations of motion, because there are no known theories for them to reduce to in the flat space limit with lower-order equations of motion. We have checked this explicitly for $k=3$.

It is worth noting that in cases that break dS invariance, either explicitly or spontaneously, higher shift symmetries can be present that act only on the spatial coordinates. Shift symmetries of this kind appear in the context of inflation~\cite{Hinterbichler:2012nm,Hinterbichler:2013dpa,Hinterbichler:2016pzn}, and have interesting consequences for correlation functions. It would be interesting to systematically classify theories where such spatial shifts appear.

\section{Conclusions} \label{sec:conclusions}
In this paper we have identified special mass values at which massive bosonic fields of all spins in (A)dS space develop shift symmetries that are the analogues of flat space polynomial shift symmetries. We have explained how these shift-symmetric fields are related to PM fields and we have constructed explicit examples of interacting scalar theories preserving the symmetries. These theories generalize many known interesting examples of shift-symmetric theories in flat space.

In flat space, shift symmetries have proven useful as an organizing principle to classify EFTs. We have taken the first steps toward such a classification in (A)dS space. In particular, we have
 considered interactions for theories with shift symmetries containing a single scalar field. 
 In addition to placing known curved-spacetime EFTs in a new context, 
we have constructed in every dimension a novel interacting scalar theory with a nonlinear quadratic shift symmetry, which is a highly nontrivial generalization of the special galileon to (A)dS space.  
Additionally, we have argued that---similar to the special galileon in flat space---this theory should have the highest possible shift symmetry while still retaining second-order equations of motion.
Therefore, the phenomenology of this theory should be extremely interesting. It possesses a potential that is completely fixed by its deformed shift symmetry, which may prove to be useful for applications in either the early or late universe.

We have focused on interactions for a single scalar field, but it is possible that there are higher-spin and/or multi-field interacting theories that are governed by deformations of the linear shift symmetries of Section \ref{linesymmsectn}.
It would be interesting to construct theories of this type.
A necessary ingredient for such a theory is a Lie algebra with symmetric or mixed-symmetry generators having up to two rows. One such algebra is the infinite higher-spin algebra underlying Vasiliev's higher-spin theory~\cite{Eastwood:2002su,Vasiliev:2003ev,Bekaert:2005vh,Joung:2014qya}. When realized as the algebra of shift symmetries, this would correspond to a putative (A)dS theory with interacting $k=0$ massive particles of every integer spin except spin one. For simpler examples, one could use the finite algebras that arise from truncations of PM higher-spin algebras~\cite{Joung:2015jza}. 
The generators of these algebras comprise all traceless Young tableaux with an even number of boxes less than or equal to $2N$, for some fixed integer $N$, and at most two rows. The $N=1$ algebra corresponds to $\mathfrak{sl}(D+1)$, which underlies the (A)dS special galileon constructed here, so an intriguing possibility is that the $N>1$ algebras are also realized by finite interacting theories of massive particles in (A)dS space.  For example, the $N=2$ algebra would correspond to an interacting theory containing spin-0 fields with $k=2$ and $k=4$, a $k=2$ spin-1 field, and a $k=0$ spin-2 field. Of course, the existence of an algebra is not by itself proof that a theory exists, so more work is required to find realizations of the algebra on fields and to find consistent invariant interactions.

Another interesting question is whether the interacting shift-symmetric theories have corresponding interacting theories of PM fields, where the algebra of shift symmetries is gauged. The (A)dS galileon corresponds to conformal gravity in this sense, and the (A)dS special galileon would correspond to a theory of a depth-0 spin-3 PM field interacting with gravity. Additionally, in this paper we have focused on bosonic fields described by symmetric tensors, but it would be interesting to extend our constructions to fermionic and mixed-symmetry fields. We also expect that the shift-symmetric fields studied here should play some interesting role in the AdS/CFT correspondence, where the bulk shift symmetries should have some novel boundary consequences.

Finally, in the same way that
shift symmetries in flat space imply enhanced soft limits of the $S$-matrix, the shift symmetries on (A)dS space considered here should imply enhanced soft limits for boundary correlation functions.  It would be interesting to explore the extent to which the (A)dS theories can be reconstructed from their soft limits.

\vspace{-.2cm}
\paragraph{Acknowledgements:} We would like to thank Lasma Alberte, Thomas Basile, Xavier Bekaert, Brando Bellazzini, Miguel Campiglia, Paolo Creminelli, Frederik Denef, Claudia de Rham, Garrett Goon, Maxim Grigoriev, Euihun Joung, Karapet Mkrtchyan, Enrico Pajer, Guilherme Pimentel, Zimo Sun, Massimo Taronna, Andrew Tolley, Sam Wong, and Michael Zlotnikov for helpful conversations and correspondence.  KH and JB acknowledge support from DOE grant DE- SC0019143 and Simons Foundation Award Number 658908.   RAR is supported by DOE grant DE-SC0011941 and Simons Foundation Award Number 555117.  AJ and RAR are supported by NASA grant NNX16AB27G.  

\appendix

\section{Symmetric polynomials}
\label{xtensorappendix}

Here we define the tensors $X^{(n)}\mn$ that are used at several points throughout the paper.  They are (A)dS versions of the tensors that appear in the decoupling limit of massive gravity \cite{deRham:2010tw,deRham:2014zqa}.

For an arbitrary symmetric tensor $M\mn$, we define the symmetric polynomials as
\be  S_n^{\rm }(M)=n!\, M^{[\mu_1}{}_{\mu_1}M^{\mu_2}{}_{\mu_2}\dots M^{\mu_n]}{}_{\mu_n} \, .\label{symmetricpolyndef}\ee
The tensors $X^{(n)}_{\mu\nu}(M)$ are defined as 
\be
 X^{(n)\mu}{}_{\nu}(M)={1\over n+1}{\delta \over \delta M^{\mu}{}_{\nu}} S_{n+1}^{\rm }(M)=(n+1)!\, \delta^{[\mu}_{\nu}M^{\mu_2}{}_{\mu_2}\dots M^{\mu_{n+1}]}{}_{\mu_{n+1}} \, .
\ee
They vanish identically for $n\geq D$.
Explicitly, the first few are given by
\begin{align} X^{(0)}_{\mu\nu}(M)&=g_{\mu\nu} \, ,\\
 X^{(1)}_{\mu\nu}(M)&=\left[M\right]g_{\mu\nu}-M_{\mu\nu} \, ,\\
  X^{(2)}_{\mu\nu}(M)&=\left(\left[M\right]^2-\left[M^2\right]\right)g_{\mu\nu}-2\left[M\right]M_{\mu\nu}+2M^2_{\mu\nu}\, , \\
   X^{(3)}_{\mu\nu}(M)&=\left(\left[M\right]^3-3\left[M\right]\left[M^2\right]+2\left[M^3\right]\right)g_{\mu\nu}-3\left(\left[M\right]^2-\left[M^2\right]\right)M_{\mu\nu}+6\left[M\right]M^2_{\mu\nu}-6M^3_{\mu\nu}\, ,
   \end{align}
where $g_{\mu \nu}$ is the background metric.

\section{Ambient space\label{embbedappendix}}

\subsection{Ambient space review}

It is often convenient to use ambient space to describe (A)dS fields \cite{Dirac:1936fq, Fronsdal:1978vb}.  To do this, we embed (A)dS$_D$ with coordinates $x^\mu$ into an ambient $(D+1)$-dimensional Minkowski space with coordinates $X^A$.  The flat ambient metric $\eta_{AB}$ has components ${\rm diag}(-1,1,1,1,\dots)$ for dS space and ${\rm diag}(-1,-1,1,1,\dots)$ for AdS space.  The (A)dS manifold is the surface defined by
\be 
\eta_{AB}X^AX^B=\pm {\cal R}^2,
\label{eq:hyperboloid}
\ee
where the constant  ${\cal R}$ is the radius.
Here and throughout, the upper sign is for dS space and the lower sign for AdS space. The (A)dS isometries are the $\mathfrak{so}(D+1)$ Lorentz transformations of the ambient space that preserve the embedded hyperboloid.
This surface can be described by the embedding functions $X^A(x)$. A basis of tangent vectors to this surface is 
\be e_\mu^A\equiv {\partial_\mu X^A}\, .\ee 
The normal vector is proportional to $X^A$, since $e_\mu^A X_A=0$.

It is often convenient to use adapted spherical coordinates $\bar A=\left(\rho,x^\mu\right)$ in the ambient space, where the metric takes the form
\be \rd s^2=\pm \rd\rho^2+\left(\rho\over {\cal R}\right)^2 \rd s^2_{ \rm (A)dS}.\ee
In these coordinates, $\rho=\sqrt{\pm X^2}$ and the (A)dS surface is at $\rho={\cal R}$.  The normal vector $X^A$ has components only along the $\rho$ direction, $X^{\bar A}=\left(\rho,0,0,0,\ldots\right)$.  The non-vanishing Christoffel symbols are
\be \Gamma^{\rho}_{\mu_1\mu_2}=\mp {\rho\over {\cal R}^2}g_{\mu_1\mu_2},\qquad \Gamma^{\mu_1}_{\mu_2\rho}={1\over \rho}\delta^{\mu_1}_{\mu_2},\qquad \Gamma^{\mu_1}_{\mu_2\mu_3}=\Gamma^{\mu_1}_{\mu_2\mu_3}\left[g\right]\,,
\ee
where $g_{\mu \nu}$ is the (A)dS metric. 

There is a bijective map between symmetric tensor fields $t_{\mu_1\cdots \mu_s}(x)$ of rank $s$ on (A)dS space and tensors fields $T_{A_1\cdots A_s}(X)$ of rank $s$ in the ambient space that satisfy the two conditions
\begin{align} 
 {\rm Homogeneity:}& \quad\left( X^A \partial_A-w \right)T_{A_1\cdots A_s}=0, \label{scalingcondine}\\
{\rm Tangentiality:}&  \quad~X^{A_1}  T_{A_1\cdots A_s}( X)=0\, ,\label{tangentialityconde}
\end{align}
where $w$ is some fixed real number called the homogeneity degree. The relation between the tensors is given by the pullback,
\be t_{\mu_1\cdots \mu_s}(x)=e^{A_1}_{\mu_1}\cdots e^{A_s}_{\mu_s}T_{A_1\cdots A_s}(X(x))\,.\ee
Any ambient space tensor can be made tangent by hitting each index with the projection operator,
\be P^{A}_{\ B}=\delta^A_{\ B}-{X^A X_B\over X^2}.\label{projectorambse}\ee

For a scalar field of homogeneity degree $w$, the field $\Phi(X)$ corresponding to $\phi(x)$ is given in the adapted spherical coordinates by
\be \Phi(\rho,x)=\left(\rho\over {\cal R}\right)^w \phi(x).\ee
For a tensor field of homogeneity degree $w$, the tangentiality condition \eqref{tangentialityconde} is the statement that the $\rho$ components of the tensor in the spherical coordinates all vanish, $T_{\rho \bar A_2\cdots \bar A_s}( X)=0$, and the condition \eqref{scalingcondine} then implies that the $\mu$ components have $\rho$ dependence given by
\be 
T_{\mu_1\cdots \mu_s}(\rho,x)=\left(\rho\over {\cal R}\right)^{w+s} T_{\mu_1\cdots \mu_s}(x).
\ee

Taking a trace of an ambient space tensor preserves the tangentiality condition \eqref{tangentialityconde} and preserves the homogeneity condition \eqref{scalingcondine} provided we lower the scaling dimension by 2, i.e. if $T_{A_1A_2\cdots A_s}$ has homogeneity degree $w$, then $\eta^{A_1A_2}T_{A_1A_2\cdots A_s}$ has homogeneity degree $w-2$.

To take derivatives in the ambient space, we first differentiate and then project all the indices using \eqref{projectorambse}.  This preserves the homogeneity degree.  There are three derivative operations on symmetric tensors, the symmetrized derivative, divergence, and Laplacian.
In~\cite{Bekaert:2010hk}, a general formula for the projection of an arbitrary number, $n$, of symmetrized ambient space partial derivatives on an arbitrary symmetric rank-$s$ tensor of homogeneity degree $w$ is given,
\be
\partial_{(A_1}\cdots\partial_{A_n}T_{A_{n+1}\cdots A_{n+s})} \leadsto \sum_{m=0}^{\lfloor \frac{n}{2}\rfloor} c_n^m \left(\pm {1\over {\cal R}^2}\right)^m g_{(\mu_1\mu_2}\cdots g_{\mu_{2m-1}\mu_{2m}}\nabla_{\mu_{2m+1}}\cdots \nabla_{{\mu_n}}t_{\mu_{n+1}\cdots\mu_{n+s})},
\ee
where a generating function for the coefficients $c_n^m$ is
\be
(1+y)^\frac{w-s}{2} \exp\left(\frac{x}{\sqrt y}\arctan\sqrt{y}\right) = \sum_{n=0}^\infty \sum_{m=0}^{\lfloor\frac{n}{2}\rfloor}\frac{1}{n!}c_n^m x^{n-2m}y^m.
\ee
This can be used to extract the $c_n^m$ in cases of interest.
The divergence involves a trace and so lowers the homogeneity degree by 2.  The corresponding (A)dS tensor is again the divergence,
\be \partial^{A_1}T_{A_{1}A_2\cdots A_{s+1}} \leadsto \nabla^{\mu_1}t_{\mu_{1}\mu_2\cdots \mu_{s+1}} \,.
\ee
The Laplacian also lowers the degree by 2, and the correspondence is
\be \square T_{A_{1}\cdots A_{s}} \leadsto \left(\nabla^2\pm{1\over {\cal R}^2}(w(D+w-1)-s) \right)t_{\mu_{1}\cdots \mu_{s}},
\ee
where $w$ is the homogeneity degree of $T_{A_{1}\cdots A_{s}}$.

\subsection{Embedding coordinates}
It will occasionally be useful to choose a particular coordinate system on the (A)dS spaces we consider. For our purposes, we will mostly be interested in the flat slicing or in stereographic coordinates.

\subsubsection{Inflationary/Poincar\'e coordinates}
We will give expressions for the dS case, the corresponding AdS expressions can be obtained straightforwardly by analytic continuation.
The inflationary slicing corresponds to the embedding\footnote{Here $\rho\in(0,\infty)$, $\eta \in (-\infty, 0)$ and $x^i\in(-\infty,\infty)$.}
\begin{subequations}
\label{eq:dsflatembedding}
\begin{align}
X^0 &= \frac{\rho}{2(-\eta)}\left(1-\eta^2+\vec x^2\right),\\
X^i &= \frac{\rho x^i}{(-\eta)},\\
X^{D} &= \frac{\rho}{2(-\eta)}\left(1+\eta^2-\vec x^2\right)~.
\end{align}
\end{subequations}
These coordinates foliate half the dS hyperboloid with flat slices, and
in these coordinates the $(D+1)$-dimensional Minkowski metric takes the form
\be
\rd s^2 = \rd \rho^2+\left( \rho H\right)^2\rd s^2_{{\rm dS}_D},
\ee
where the dS metric in the inflationary slicing is given by
\be
\rd s^2_{{\rm dS}_D} = \frac{1}{H^2\eta^2}\left(-\rd\eta^2 +\rd\vec x^2\right).
\ee
It is straightforward to check that the coordinates~\eqref{eq:dsflatembedding} satisfy~\eqref{eq:hyperboloid} with $ {\cal R}= H^{-1} $, so that the slice $\rho=H^{-1}$ is the dS hyperboloid with Hubble parameter $H$.
It is often useful to construct the following ambient lightcone coordinates:
\begin{align}
\label{eq:xpluslightcone}
X^+ &= X^0+X^D=\frac{\rho }{(-\eta)},\\
X^- &= X^0-X^D =\frac{\rho}{(-\eta)}\left(-\eta^2+\vec x^2\right) .
\end{align}
By inverting the embedding~\eqref{eq:dsflatembedding}, we can express the ambient space derivatives in terms of intrinsic coordinates~\cite{Goon:2011qf}
\begin{subequations}
\label{eq:dsflatderivs}
\begin{align}
\frac{\partial}{\partial X^0} &= \frac{1}{2(-\eta)}\left(-1+\eta^2 -\vec x^2\right)\partial_\rho+\frac{1}{2\rho}(1+\eta^2+\vec x^2)\partial_\eta +\frac{\eta}{\rho}x^i\partial_i,\\
\frac{\partial}{\partial X^i} &= \frac{x^i}{(-\eta)}\partial_\rho-\frac{x^i}{\rho}\partial_\eta-\frac{\eta}{\rho}\partial_i,\\
\frac{\partial}{\partial X^D} &= \frac{1}{2(-\eta)}\left(1+\eta^2 -\vec x^2\right)\partial_\rho+\frac{1}{2\rho}(-1+\eta^2+\vec x^2)\partial_\eta +\frac{\eta}{\rho}x^i\partial_i.
\end{align}
\end{subequations}
The equations~\eqref{eq:dsflatembedding} and~\eqref{eq:dsflatderivs}
can then be used to translate ambient space formulae into (A)dS formula in the flat slicing.

\subsubsection{Stereographic coordinates}
\label{sec:sterocoords}
Another coordinate system which is often useful (particularly in Appendix~\ref{cosetappendix}), corresponds to the embedding\footnote{Here the coordinates $\rho\in(0,\infty)$ and $x^\mu\in (-\infty, \infty)$.}
\begin{subequations}
\label{eq:stereoembedding}
\begin{align}
X^\mu &= x^\mu \frac{\rho H}{1+\frac{H^2}{4}x^2}\\
X^{5} &= \rho \frac{ 1-\frac{H^2}{4}x^2 }{1+\frac{H^2}{4}x^2},
\end{align}
\end{subequations}
so that the ambient Minkowski metric takes the form
\be
\rd s^2 = \rd \rho^2+\left( \rho H\right)^2\left(\frac{1}{1+\frac{H^2}{4}x^2}\right)^2\eta_{\mu\nu}\rd x^\mu\rd x^\nu,
\ee
where again the $\rho = H^{-1}$ slice corresponds to a dS space with Hubble parameter $H$. These coordinates cover $3/4$ of the dS hyperboloid. It is  straightforward to obtain AdS coordinates by analytically continuing $H^2 \mapsto -L^{-2}$.

In this case also we can invert the coordinate embedding~\eqref{eq:stereoembedding} to express ambient space derivatives as~\cite{Biswas:2002nk}
\begin{subequations}
\begin{align}
\frac{\partial}{\partial X^\mu} &= \frac{1}{\rho H} \left(\delta_\mu^\nu\left(1+\frac{H^2}{4}x^2\right)-\frac{H^2}{2}x^\nu x_\mu\right)\partial_\nu+\frac{H x_\mu}{1+\frac{H^2}{4}x^2}\partial_\rho,\\
\frac{\partial}{\partial X^D} &= -\frac{x^\mu}{\rho}\partial_\mu+\left(\frac{ 1-\frac{H^2}{4}x^2 }{1+\frac{H^2}{4}x^2}\right)\partial_\rho.
\end{align}
\end{subequations}

\section{Coset constructions\label{cosetappendix}}
Here we describe the coset construction of the abelian scalar interactions described in Section~\ref{sec:interactingscalars}. The advantage of this approach is that it is entirely systematic. Once we have classified the possible algebras, we can algorithmically construct a scalar field theory that nonlinearly realizes the breaking of this algebra down to the (A)dS algebra; see, {\it e.g.},~\cite{Goon:2012dy} for an introduction to the formalism.

\subsection{(A)dS galileons}
We first consider the interactions for a $k=1$ scalar with undeformed algebra, {\it i.e.},  the (A)dS galileons \cite{Goon:2011uw,Burrage:2011bt,Goon:2011qf}. We will find that, unlike the flat case, only one of the (A)dS galileons is a Wess--Zumino term.  
The coset construction of the scalar theory that nonlinearly realizes the deformed $k=1$ algebra, $\mathfrak{so}(D+2)$, has been considered in~\cite{Clark:2005ht,Hinterbichler:2012mv}.

As discussed in Section \ref{algebrassection}, the $k=1$ scalar is invariant under the (A)dS isometries, which form an $\frak{so}(D,1)$ algebra, 
along with the following nonlinearly realized symmetry written in ambient space:
\be
\delta \Phi = S_AX^A.
\ee
Taken together, these symmetries close to form the algebra
\begin{align}
[J_{AB},J_{CD}] &= \eta_{AC}J_{BD}-\eta_{BC}J_{AD}+\eta_{BD}J_{AC}-\eta_{AD}J_{BC},\\
[J_{AC}, S_D] &= \eta_{AD}S_C-\eta_{CD}S_A,
\end{align}
which is isomorphic to the $(D+1)$-dimensional Poincar\'e algebra, $\frak{iso}(D,1)$.\footnote{This is the same algebra nonlinearly realized by the flat space DBI action in $D$ dimensions. Here we are linearly realizing a different subalgebra.} 
We can write this algebra in $D$-dimensional language by defining 
\begin{align}
J_{(D+1)\mu} &\equiv H^{-1}\hat P_\mu,& J_{\mu\nu}&\equiv \hat J_{\mu\nu},\\
S_{(D+1)} &\equiv - H^{-1}C, & S_\mu&\equiv B_\mu,
\end{align}
after which the commutation relations take the form
\begin{subequations}
\begin{align}
[\hat P_\mu, \hat P_\nu] &= H^2 \hat J_{\mu\nu}\, , \qquad
[\hat P_\mu, B_\nu] = \eta_{\mu\nu}C \,, \qquad [\hat P_\mu, C] = -H^2 B_\mu \, ,\\
[\hat J_{\mu\nu}, \hat P_\rho] &= \eta_{\mu\rho}\hat P_\nu-\eta_{\nu\rho}\hat P_\mu\, , \quad [\hat J_{\mu\nu}, B_\rho] = \eta_{\mu\rho}B_\nu-\eta_{\nu\rho}B_\mu \,, \\
[\hat J_{\mu\nu}, \hat J_{\rho\sigma}] &= \eta_{\mu\rho}\hat J_{\nu\sigma}-\eta_{\nu\rho}\hat J_{\mu\sigma}+\eta_{\nu\sigma}\hat J_{\mu\rho}-\eta_{\mu\sigma}\hat J_{\nu\rho} \, .
\end{align}
\label{eq:dsgalalg}
\end{subequations}
In the $H\to 0$ limit, this reduces to the flat space galileon algebra studied in Ref.~\cite{Goon:2012dy}. 

\subsubsection*{Maurer--Cartan form} 
We are considering the symmetry-breaking pattern
\be
\frak{iso}(D,1)\longrightarrow \frak{so}(D,1),
\ee
which linearly realizes the dS subalgebra. The coset space is then parametrized by $D$ coordinates $y^\mu$, and $D+1$ Goldstone fields $\xi^\mu$, $\phi$, as
\be
g = e^{y^\mu\hat P_\mu}e^{\xi^\mu B_\mu}e^{\phi C} \, .
\label{eq:k1cosetparam}
\ee
In order to construct invariant Lagrangians, we require the Maurer--Cartan form, $\omega = g^{-1}\rd g$. This can be computed directly,
\be
\omega = \rd y^\mu e_\mu^\nu \hat P_\nu+\left(\rd\xi^\nu+2\xi_\mu \omega_{\rm spin}^{\mu\nu}-\phi H^2 \rd y^\mu e_\mu^\nu\right) B_\nu +\left(\rd\phi+\xi_\nu e_\mu^\nu \rd y^\mu\right)C+\omega_{\rm spin}^{\mu\nu}\hat J_{\mu\nu} \, ,
\ee
where $e_\mu^\nu$ and $\omega_{\rm spin}^{\mu \nu}$ are a vielbein and spin connection for dS space
\begin{align}
e_\mu^\nu &=  \left(\delta_\mu^\nu-\frac{y_\mu y^\nu}{y^2}\right) \frac{\sin\sqrt{H^2y^2}}{\sqrt{H^2y^2}}+\frac{y_\mu y^\nu}{y^2},\\
\omega_{\rm spin}^{\mu\nu} &= \left(1-\cos\sqrt{H^2y^2}\right)\left[\frac{y^\mu\rd y^\nu-y^\nu\rd y^\mu}{2y^2}\right].
\end{align}
This can be made manifest by the coordinate redefinition~\cite{Clark:2005ht}
\be
y^\mu = x^\mu \sqrt\frac{4}{H^2 x^2}\arctan \sqrt\frac{H^2x^2}{4}, 
\label{eq:projectivecoords}
\ee
after which the vielbein takes the stereographic form discussed in Section~\ref{sec:sterocoords},
\be
e_\mu^\nu = \left(\frac{1}{1+\frac{H^2}{4} x^2}\right)\delta_\mu^\nu.
\label{eq:projectiveviel}
\ee
Additionally, note that the covariant exterior derivative acts on a field $\psi$ as
\be
\rd^{(\nabla)} \psi \equiv \rd\psi+\rho(\hat J)\psi,
\ee
where $\rho(\hat J)$ is the matrix representation of the spin generators in the representation under which $\psi$ transforms. In the vector case, using
\be
\omega^{\mu\nu}_{\rm spin}\hat J_{\mu\nu} \xi_\alpha = \omega^{\mu\nu}_{\rm spin}\left(\xi_\mu\eta_{\nu\alpha} - \xi_\nu\eta_{\mu\alpha}\right) = 2\xi_\mu \omega^{\mu\alpha}_{\rm spin},
\ee
we see that the combination appearing in the Maurer--Cartan form is exactly the covariant exterior derivative of $\xi$. 
We can make two simplifications:  the vielbein factors serve to convert indices to dS indices, and we can use the fact that everything is dS covariant 
to work in an arbitrary coordinate system.
The dS covariant Maurer--Cartan components can then be written as
\begin{align}
\omega_{\hat P}^\mu &= e_\alpha^\mu\rd x^\alpha\, ,\\
\omega_B^\mu &= \rd^{(\nabla)} \xi^\mu - H^2\phi \rd x^\mu\, ,\\
\omega_C &= \rd\phi +\xi_\mu \rd x^\mu\,.
\end{align}
We are interested in constructing invariant actions for the scalar field, $\phi$; we would therefore like to remove the redundant vector Goldstone, $\xi_\mu$.
That this is possible follows from the commutator $[\hat P , B] \propto C$, which implies that we can remove $\xi_\mu$ by setting $\omega_C = 0$. Doing this, we find the inverse Higgs constraint\footnote{Since $[\hat P, C] \sim B$, there appears to be another inverse Higgs constraint we could impose which would eliminate the scalar Goldstone and leave a theory of a vector nonlinearly realizing the (A)dS galileon symmetry.
The inverse Higgs constraint $\omega_B = 0$ eliminates $\phi$ in favor of the trace of $\nabla_\mu \xi_\alpha$ as $\phi = \frac{1}{DH^2}\nabla\cdot\xi$, so that the building blocks for invariant actions are
\begin{align}
\omega_C = \rd x^\mu\left( \frac{1}{DH^2}\nabla_\mu\nabla\cdot\xi +\xi_\mu\right)\, ,\\
\omega_B^\mu = \rd x^\nu\left( \nabla_{(\mu}\xi_{\nu)_T}+ \nabla_{[\mu}\xi_{\nu]}\right).
\end{align}
It would be interesting to investigate this further.}
\be
\xi_\mu = -\nabla_\mu \phi.
\ee
Substituting this constraint into $\omega_B$, the remaining invariant building block is 
\be
\omega_B^\mu = -\rd x^\nu\left(\nabla^\mu\nabla_\nu +\delta_\nu^\mu H^2\right)\phi = -\rd x^\nu \Phi^{(1)}_\nu{}^\mu,
\ee
where we have defined
\be
\Phi^{(1)}_{\mu\nu} = \left(\nabla_\mu\nabla_\nu+H^2g_{\mu\nu}\right)\phi \, ,
\ee
which is the same invariant tensor \eqref{eq:k1invariant} that is natural from ambient space perspective.
\subsubsection*{Interactions and Wess--Zumino terms}
After imposing the inverse Higgs constraint to remove the vector Goldstone, the only remaining building blocks are $\omega_{\hat P}^\mu$ and $\omega_B^\mu$. The simplest interaction terms we can construct from these ingredients are $D$-forms:
\begin{subequations}
\label{eq:Dform1}
\begin{align}
\omega_1&= \epsilon_{\mu_1\mu_2\cdots\mu_D}\omega_B^{\mu_1}\wedge\omega_{\hat P}^{\mu_2}\wedge\omega_{\hat P}^{\mu_3}\wedge\cdots\wedge\omega_{\hat P}^{\mu_D} ,\\
\omega_2&= \epsilon_{\mu_1\mu_2\cdots\mu_D}\omega_B^{\mu_1}\wedge\omega_{B}^{\mu_2}\wedge\omega_{\hat P}^{\mu_3}\wedge\cdots\wedge\omega_{\hat P}^{\mu_D},\\
&~~\vdots\\
\omega_D&= \epsilon_{\mu_1\mu_2\cdots\mu_D}\omega_B^{\mu_1}\wedge\omega_{B}^{\mu_2}\wedge\omega_{B}^{\mu_3}\wedge\cdots\wedge\omega_{B}^{\mu_D},
\end{align}
\end{subequations}
where $\omega_n$ contains $n$ factors of $\omega_B^{\mu}$.
In terms of the tensor $\Phi^{(1)}_{\mu\nu}$, these give the Lagrangians
\begin{subequations}
\label{eq:4AdSGals}
\begin{align}
{\cal L}_1 & =- \sqrt{-g}\epsilon^{\mu \alpha_2\cdots\alpha_D}\epsilon_{\nu\alpha_2\cdots\alpha_D} \Phi^{(1)}_\mu{}^\nu ,\\
{\cal L}_2 &= \sqrt{-g}\epsilon^{\mu_1 \mu_2 \alpha_3\cdots\alpha_D}\epsilon_{\nu_1\nu_2\alpha_3\cdots\alpha_D} \Phi^{(1)}_{\mu_1}{}^{\nu_1}\Phi^{(1)}_{\mu_2}{}^{\nu_2}  ,\\
&~~\vdots\\
{\cal L}_n &=(-1)^n\sqrt{-g}\epsilon^{\mu_1 \mu_2\cdots\mu_n\alpha_{n+1}\cdots\alpha_D}\epsilon_{\nu_1\nu_2\cdots\nu_n\alpha_{n+1}\cdots\nu_D}\Phi^{(1)}_{\mu_1}{}^{\nu_1}\Phi^{(1)}_{\mu_2}{}^{\nu_2}\cdots\Phi^{(1)}_{\mu_n}{}^{\nu_n}, \\
&~~\vdots\\
{\cal L}_D &=(-1)^D\sqrt{-g}\epsilon^{\mu_1 \mu_2\cdots\mu_D}\epsilon_{\nu_1\nu_2\cdots\nu_D}\Phi^{(1)}_{\mu_1}{}^{\nu_1}\Phi^{(1)}_{\mu_2}{}^{\nu_2}\cdots\Phi^{(1)}_{\mu_D}{}^{\nu_D} \, .
\end{align}
\end{subequations}
Though it is not obvious, for $D=4$ these Lagrangians are equivalent up to total derivatives to the first four of the (A)dS galileons presented in \cite{Goon:2011uw,Burrage:2011bt,Goon:2011qf}.  The presentation here also appears in~\cite{deRham:2018svs}.

In flat space, each of the galileon interactions is a Wess--Zumino term, but here they are not.\footnote{We define Wess--Zumino terms as elements of the relative Lie algebra cohomology between $\frak{iso}(D,1)$ and $\frak{so}(D,1)$. See~\cite{Goon:2012dy} for more details.}   The difference is due to the different symmetry algebras in the (A)dS space and flat space theories. In flat space, the invariant $(D+1)$-forms that give rise to the galileon Lagrangians are
\be
\tilde \omega_n= \epsilon_{\mu_1\mu_2\cdots\mu_D}\omega_C\wedge\omega_B^{\mu_1}\wedge\cdots\wedge\omega_{B}^{\mu_{n-1}}\wedge\omega_{\hat P}^{\mu_{n}}\wedge\cdots\wedge\omega_{\hat P}^{\mu_D}.
\label{eq:D1form1}
\ee
These forms are closed in both flat space and (A)dS space, but only in (A)dS space can they be written as the exterior derivative of an invariant $D$-form. 
To see this, note that the Maurer--Cartan forms satisfy the Cartan structure equations
\be
{\rm d}\omega^{i}=-\frac{1}{2}f_{jk}{}^{i}\omega^{j}\wedge\omega^{k},
\ee
where $f_{ij}^{\ \ k}$ are the Lie algebra structure constants.\footnote{Explicitly, given a basis, $\frak{e}_i$, of the Lie algebra, the structure constants are defined by
\be 
[\frak{e}_i,\frak{e}_j]=f_{ij}^{\ \ k}\frak{e}_k \ .
\ee
}
This defines the action of the exterior derivative on the forms.
Since $C$ commutes with everything in the $H\to 0$ limit of the algebra~\eqref{eq:dsgalalg}, the forms \eqref{eq:D1form1} cannot be written as the exterior derivative of a $D$-form in flat space.
However, in curved space we have the nonzero commutator
\be
 [\hat P_\mu, C] = -H^2 B_\mu,
\ee
which implies that
\be
\rd \omega_B^\mu = \frac{H^2}{2}\omega_{\hat P}^\mu\wedge\omega_C.
\ee
This means that, when $H \neq 0$, the forms~\eqref{eq:D1form1} can be written as the exterior derivative of the $D$-forms~\eqref{eq:Dform1}, so they are not Wess--Zumino terms. 

Note that ${\cal L}_{D+1}$ is missing from the list \eqref{eq:4AdSGals}.  This interaction cannot be written solely using $\Phi^{(1)}_{\mu\nu}$. It is a Wess--Zumino term in (A)dS space, derived from the $(D+1)$-form
\be
\omega_{D+1}= \epsilon_{\mu_1\mu_2\cdots\mu_D}\omega_C\wedge\omega_B^{\mu_1}\wedge\omega_{B}^{\mu_{2}}\wedge\cdots\wedge\omega_{B}^{\mu_D},
\ee
which is closed but cannot be written as the exterior derivative of an invariant 4-form. This is because $C$ commutes with $B$ even in (A)dS space.
Pulling this object back to the physical $D$-dimensional space, it leads to the Lagrangian
\be
{\cal L}_{D+1} = (-1)^D\sqrt{-g}\epsilon^{\mu_1 \mu_2 \cdots\mu_D}\epsilon_{\nu_1\nu_2\cdots\nu_D}\phi \Phi^{(1)}_{\mu_1}{}^{\nu_1}\Phi^{(1)}_{\mu_2}{}^{\nu_2}\cdots\Phi^{(1)}_{\mu_D}{}^{\nu_D},
\label{eq:5AdSGal}
\ee
which cannot be written solely in terms of $\Phi^{(1)}_{\mu\nu}$.

\subsubsection*{Field transformations}
We now work out the transformation rules for the Goldstone fields. We do this by acting on the coset parametrization~\eqref{eq:k1cosetparam} with the transformation
\be
g' = e^{\alpha C} \quad {\rm or} \quad g'=e^{\beta^\mu \hat B_\mu},
\ee
and then combining the exponentials using Baker--Campbell--Hausdorff formulas.
We only need to work to first order in the parameters $\alpha$ and $\beta$, which will give us the infinitesimal form of the symmetry we are interested in.
Going to projective coordinates~\eqref{eq:projectivecoords}, the field transformations take the form
\begin{subequations}
\label{eq:cosettransforms}
\begin{align}
\delta_{C}\phi &= -1+\frac{2}{1+\frac{H^2}{4}x^2}, &  \delta_{C}\xi^\mu &= \frac{H^2x^\mu}{1+\frac{H^2}{4}x^2},\\
\delta_{ B_\mu}\phi &= -\frac{x^\mu}{1+\frac{H^2}{4}x^2},  &  \delta_{ B^\mu}\xi^\nu &= \eta^{\mu\nu}-\frac{H^2 x^\mu x^\nu}{2\left(1+\frac{H^2}{4}x^2\right)}.
\end{align}
\end{subequations}
Recalling that the stereographic embedding is given by~\eqref{eq:stereoembedding}, we see that 
\be
\delta_C\phi \propto X^5, \qquad\quad \delta_{B^\mu}\phi \propto X^\mu,
\ee
as anticipated. We can also check that the transformation rules~\eqref{eq:cosettransforms} are
compatible with the inverse Higgs constraints.

\subsection{$k=2$ scalars}
We now discuss the $k=2$ scalars from the coset perspective. In this case, the coset construction of the non-abelian theory can be performed, but the resulting objects are rather complicated, so we focus on the abelian theory which realizes the symmetries of the linear theory.

For completeness, we split the deformed $k=2$ algebra into (A)dS-covariant form. Starting with
\begin{align}
[J_{AB},J_{CD}] &= \eta_{AC}J_{BD}-\eta_{BC}J_{AD}+\eta_{BD}J_{AC}-\eta_{AD}J_{BC},\\
[J_{AC}, S_{CD}] &= \eta_{AC}S_{BD}-\eta_{BC}S_{AD}+\eta_{AD}S_{BC}-\eta_{DB}S_{AC},\\
[S_{AB}, S_{CD}] &= \alpha\left(\eta_{AC}J_{BD}+\eta_{BC}J_{AD}+\eta_{BD}J_{AC}+\eta_{AD}J_{BC}\right),
\end{align}
and defining the following $D$-dimensional generators (the normalizations are chosen to reproduce the algebra studied in~\cite{Hinterbichler:2015pqa} in the $H\to 0$ limit),
\begin{align}
J_{(D+1)\mu} &\equiv \hat H^{-1}\hat P_\mu, & S_{(D+1)(D+1)} &\equiv \frac{D}{D+1}H^{-1}\hat C,\\
S_{(D+1)\mu} &\equiv -\hat B_\mu, & S_{\mu\nu} &\equiv \hat S_{(\mu\nu)_T}-\frac{1}{D+1}H^{-1}\eta_{\mu\nu}\hat C,
\end{align}
we can write this algebra in $D$-dimensional language,
\begin{align}
[\hat P_\mu, C] &= -\frac{2(D+1)}{D}H^2\hat B_\mu,~~~~~~~~[\hat P_\mu, \hat B_\nu] = \eta_{\mu\nu}C-H^2\hat S_{\mu\nu},\\
[\hat P_\mu, \hat S_{\nu\rho}] &= \eta_{\mu\nu}\hat B_\rho+\eta_{\mu\rho}\hat B_\nu-\frac{2}{D}\eta_{\nu\rho}\hat B_\mu,~~~~~~~~[J_{\mu\nu},\hat B_\rho] = \eta_{\mu\rho}\hat B_\nu-\eta_{\nu\rho}\hat B_\mu,\\
[J_{\mu\nu},\hat S_{\rho\sigma}] &= \eta_{\mu\rho}\hat S_{\nu\sigma}-\eta_{\nu\rho}\hat S_{\mu\sigma}+\eta_{\mu\sigma}\hat S_{\nu\rho}-\eta_{\nu\sigma}\hat S_{\mu\rho},\\\nonumber
[\hat B_\mu, C] &= \alpha\frac{2(D+1)}{D}H^2 \hat P_\mu,~~~~~~~~[\hat B_\mu, \hat B_\nu] = \alpha H^2 J_{\mu\nu},\\
[\hat B_\mu, \hat S_{\nu\rho}] &= -\alpha \left(\eta_{\mu\nu}\hat P_\rho+\eta_{\mu\rho}\hat P_\nu-2\eta_{\nu\rho}\hat P_\mu/D\right),\\
[\hat S_{\mu\nu}, \hat S_{\rho\sigma}] &= \alpha\left(\eta_{\mu\rho}J_{\nu\sigma}+\eta_{\nu\rho}J_{\mu\sigma}+\eta_{\nu\sigma}J_{\mu\rho}+\eta_{\mu\sigma}J_{\nu\rho}\right),
\end{align}
and with the commutators between $P_\mu$ and $J_{\mu\nu}$ the same as in \eqref{eq:dsgalalg}.
The coset construction proceeds by nonlinearly-realizing the symmetries generated by $\hat C, \hat B_\mu, \hat S_{\mu\nu}$. We can accomplish this by parametrizing the coset as
\be
g = e^{y\cdot\hat P}e^{\phi\hat C}e^{\xi\cdot\hat B}e^{A\cdot\hat S}.
\ee
Ultimately, we want an action for only the scalar degree of freedom; we see from the algebra that there are possible inverse Higgs constraints which eliminate the vector, $\xi_\mu$, and symmetric traceless tensor, $A_{\mu\nu}$, in terms of $\phi$.

\subsubsection*{Abelian Maurer--Cartan form}
We focus on the abelian theory, where the symmetries act on the field as
\be
\delta\Phi = S_{AB}X^AX^B.
\label{eq:k2linearsymmetry}
\ee
The Maurer--Cartan form is given by:
\begin{align}
\nonumber
\omega =&~ 
\rd x^\mu e_\mu^\alpha\hat P_\alpha+\left(\rd\phi+\xi_\mu\rd x^\mu \right)\hat C+ \left(\rd^{(\nabla)} \xi^\mu +2\rd x^\nu  A^\mu_\nu-2H^2\left(\frac{D+1}{D}\right)\phi\,\rd x^\mu \right)\hat B_\mu\\
&+\left(\rd^{(\nabla)} A^{\mu\nu}-H^2\rd x^\mu  \xi^\nu\right)\hat S_{\mu\nu},
\label{eq:MCformend}
\end{align}
where all indices are raised and lowered with the dS metric and we have changed to stereographic coordinates as in the $k=1$ case. Everything is covariant, so we can work in an arbitrary slicing.

There are many possible ways of imposing inverse Higgs constraints, but we are interested in scalar actions which nonlinearly realize the relevant symmetries, so we want to remove $\xi_\mu$ and $A_{\mu\nu}$ in favor of $\phi$. This can be achieved by imposing the inverse Higgs constraints
\be
\omega_{\hat B}^\mu = \omega_{\hat C} = 0.
\ee
In solving for $A_{\mu\nu}$, we have to take into account that it is traceless, so we only have to set the traceless part of $\omega_{\hat B}^\mu$ to zero.
Solving both of these constraints leads to
\be
\xi_\mu = -\nabla_\mu\phi, \qquad\qquad A_{\mu\nu} = \frac{1}{2}\nabla_{(\mu}\nabla_{\nu)_T}\phi.
\ee
Inserting this back into the Maurer--Cartan form, we find that the remaining building blocks are
\begin{align}
\omega_{\hat P}^\mu &= e_\alpha^\mu\rd x^\alpha,\\
\omega_{\hat B}^\mu &= -\rd x^\mu\frac{1}{D}\left(\square+2(D+1)H^2\right)\phi,\\
\omega_{\hat S}^{\mu\nu} &= \rd x^\alpha\left(\frac{1}{2} \nabla_\alpha \nabla^{(\mu}\nabla^{\nu)_T}\phi +H^2 \delta_\alpha^{(\mu} \nabla^{\nu)_T} \phi\right),
\end{align}
where all indices should be raised and lowered with the dS metric. Note that the $\omega_B$ building block is the $k=2$ linear equation of motion, which is exactly invariant under the symmetries. The 3-index tensor appearing in $\omega_S$ looks superficially different from~\eqref{eq:phi2tensor}, but we can write
\be
\frac{1}{2} \nabla_\alpha \nabla_{(\mu}\nabla_{\nu)_T}\phi +H^2 g_{\alpha (\mu} \nabla_{\nu)_T} \phi = \frac{1}{2}\Phi^{(2)}_{\alpha\mu\nu} -\frac{1}{2D}g_{\mu\nu}\nabla_\alpha\left(\square+2(D+1)H^2\right)\phi.
\ee
So we see that we can use the invariant objects
\begin{align}
\Phi_{\mu\nu\rho}^{(2)} &=  \left(\nabla_{(\mu}\nabla_\nu\nabla_{\rho)}+4H^2 g_{(\mu\nu}\nabla_{\rho)}\right)\phi,\\
\Phi^{(2)} &= \left(\square+2(D+1)H^2\right)\phi,
\end{align}
to construct actions exactly invariant under the symmetries~\eqref{eq:k2linearsymmetry}. It is possible that there are ghostly Wess--Zumino terms with fewer derivatives than naively expected and which are invariant under the linear symmetries only up to a total derivative. We have not searched for them systematically, but it would be interesting to classify these for all of the abelian shift-symmetric scalars.

\section{Higher-order conformal scalars}
\label{app:GJMSscalar}

In this appendix we discuss a relation between the shift-symmetric scalar theories~\eqref{massivegenadse} and higher-order conformal scalars in even dimensions.

Consider a free scalar on flat space, with the $2n$\textsuperscript{th} order Lagrangian ${\cal L}\sim \phi\square^n\phi$.  This is conformally invariant on flat space, and the field $\phi$ has conformal dimension $\Delta={D\over 2}-n$.  We would like to conformally couple this CFT to a background metric $g_{\mu\nu}$ so that the resulting theory is Weyl invariant with the appropriate conformal weight,
\be g_{\mu\nu}\mapsto e^{2\sigma(x)}g_{\mu\nu},\qquad \phi\mapsto e^{-{\Delta}\sigma(x)}\phi,\qquad\Delta={D\over 2}-n,\ee
for an arbitrary scalar function $\sigma(x)$.  

This conformal coupling to the metric can be achieved by replacing the free Klein--Gordon operator $\square$ with the covariant one and then adding specific lower-derivative terms proportional to the background curvatures,
\be  \int \rd^Dx \,  \phi \square^n \phi\rightarrow  \int \rd^Dx \sqrt{\lvert g\rvert}\left( \phi  \square^n\phi+{\rm curvature\ terms}\right),\ee
 in such a way that the action is Weyl invariant.  This Weyl invariance can be achieved in all cases except the low even dimensions $D=2,4,\ldots,2n-2$.  With the right choice of curvature terms, the scalar equations of motion give conformally covariant operators known as the GJMS operators \cite{Graham01121992}, 
$ P_n[g]$,
which are operators of order $2n$ that reduce to $\square^n$ in flat space
and which are covariant under the Weyl transformations
\be P_n[e^{2\sigma} g]\left(e^{-{\Delta}\sigma}\phi\right)= e^{{-(\Delta+2n)}\sigma}  P_n[ g]\phi  .\ee
For $D=2,4,\ldots,2n-2$, the curvature terms become singular, so the GJMS operators do not exist and the CFT cannot be coupled to a metric in a Weyl-invariant manner \cite{Karananas:2015ioa,Farnsworth:2017tbz}.

The general expressions for the curvature terms are complicated \cite{2009arXiv0905.3992J,Juhl:2011ua,2012arXiv1203.0360F}, but in the case where the background metric is maximally symmetric, the action can be written in a simple factorized form \cite{Beccaria:2015vaa} 
\be \int \rd^Dx \,  \phi\, \square^n\phi\rightarrow  \int \rd^D x \sqrt{-g}\, \phi  \prod_{l=1}^n \left[ \square-H^2\left({D\over 2}-l\right)\left({D\over 2}+l-1\right)\right]\phi .\label{generaleinsts}\ee
Some general properties are discussed in \cite{Gover:2005mn,Manvelyan:2006bk,Manvelyan:2007tk,2009arXiv0905.3992J,Juhl:2011ua,2012arXiv1203.0360F}.  In low even dimensions $D=2,4,\ldots, 2n-2$, the action \eqref{generaleinsts} is non-singular.  Even though the general expression is singular in these dimensions, the singularities cancel upon using the maximal symmetry condition.

We see from the factorized form that when $H^2\not=0$ the action \eqref{generaleinsts} describes $n$ propagating scalars with masses
\be 
m^2_l=H^2\left({D\over 2}-l\right)\left({D\over 2}+l-1\right).
\ee
Comparing to \eqref{scalarmassvaluese}, we see that in $D=2$ this contains the first $n$ of the shift-symmetric scalars.  In $D=4$, it contains the first $n-1$ of the shift-symmetric scalars, along with the conformal scalar with mass $m^2=2H^2$.  For $D\geq 6$ with $D$ even, we get the first $n+1-D/2$ shift-symmetric scalars along with the conformal scalar and more massive scalars.  For $D$ odd, the spectra do not contain any of the shift-symmetric scalars. This connection in $D=4$ was noted in Appendix A.2 of \cite{Baumann:2017jvh}.

\section{Massive spin-3 decoupling limits}\label{spin3Appendix} 
We have seen that shift-symmetric fields appear as longitudinal components of higher-spin fields at PM points. 
The massive spin-2 case was discussed in Section~\ref{PMsection}.
In this appendix, we discuss the next simplest case of a free massive spin-3 field.
This case has two novelties compared to the massive spin-2 case: there are two PM decoupling limits, and for one of these the longitudinal mode is a vector, instead of a scalar. A free massive spin-3 field in (A)dS space is described by the Lagrangian
\begin{align} 
{\cal L}^{s=3}_{m^2}=&-{1\over 2}\nabla_\mu b_{\nu\alpha\beta} \nabla^\mu b^{\nu\alpha\beta}+{3\over 2}\nabla_\mu b^{\mu}_{\ \alpha\beta} \nabla_\nu b^{\nu\alpha\beta}-3\nabla_\mu b_{\nu\alpha}^{\ \ \alpha} \nabla_\beta b^{\beta \mu\nu} +{3\over 2} \nabla_\mu b_{\nu\alpha}^{\ \ \alpha}\nabla^\mu b^{\nu\beta}_{\ \ \beta}+{3\over 4}\nabla_\mu b^{\mu\alpha}_{\ \ \alpha} \nabla_\nu b^{\nu\beta}_{\ \ \beta} \nonumber\\\nonumber
&-{1\over 2}m^2\left(b_{\mu\nu\alpha}b^{\mu\nu\alpha}-3b_{\mu\alpha}^{\ \ \alpha}b^{\mu\beta}_{\ \ \beta}\right)-{3(D-2)\over 2D}m h\nabla_\mu b^{\mu\nu}_{\ \ \nu}+{3(D-2)(D-1)\over 2D^2}(\nabla h)^2+{9\over 4}m^2 h^2\\
&+\frac{(D-1)H^2}{2}\left(\frac{D-3}{D-1}b_{\mu\nu\rho}b^{\mu\nu\rho} - 6 b_{\mu\alpha}^{~~\alpha}b^{\mu\beta}_{~~\beta}-9h^2\right),
\label{eq:spin3curvedspacelag}
\end{align}
where $b_{\mu\nu\rho}$ is a symmetric tensor describing the spin-3 particle and $h$ is an auxiliary scalar field. The auxiliary field vanishes on shell, but it is needed to impose the constraints that $b_{\mu\nu\rho}$ is transverse and traceless. This Lagrangian can be obtained, for example, by radial dimensional reduction~\cite{Biswas:2002nk,Hallowell:2005np}.

There are three special masses for which the Lagrangian \eqref{eq:spin3curvedspacelag} develops a gauge symmetry, corresponding to the different PM depths, $t$:
\begin{itemize}

\item $t=2$, $m=0$: This is the massless theory. At this point, the auxiliary scalar $h$ completely decouples from $b_{\mu\nu\rho}$ and the action for $b_{\mu\nu\rho}$ is invariant under the gauge symmetry
\be
\delta b_{\mu\nu\rho} = \nabla_{(\mu}\Lambda_{\nu\rho)} \, ,
\label{freegauges}
\ee
where the gauge parameter $\Lambda_{\mu \nu}$ is symmetric and traceless.  The action for $b_{\mu\nu\rho}$ is therefore that of a Fronsdal field and it propagates only the helicity-3 modes.  The gauge symmetry removes the $0$, $\pm1$, and $\pm2$ helicity modes.

\item $t=1$, $m^2 = D H^2$: At this PM point the action develops a gauge symmetry with a vector gauge parameter, $\xi_\mu$,\footnote{This gauge symmetry involves a square root of $H^2$, so to keep the Lagrangian real we must replace $h\rightarrow ih$ when we send $H^2\rightarrow -L^{-2}$.}
\begin{align}
\label{t1PMgauge1}
\delta b_{\mu\nu\rho} &= \nabla_{(\mu}\nabla_{\nu}\xi_{\rho)} - \frac{1}{D}g_{(\mu\nu}\nabla_{\rho)}\nabla_\alpha\xi^\alpha+ H^2 g_{(\mu\nu}\xi_{\rho)}\, ,\\
\delta h &= -\frac{1}{3} \sqrt{DH^2} \nabla_\mu\xi^\mu~.
\label{t1PMgauge2}
\end{align}
This gauge symmetry removes the $0$ and $\pm 1$ helicity components of $b_{\mu\nu\rho}$, leaving propagating $\pm2$, and $\pm3$ helicities.

\item $t=0$, $m^2 = 2(D-1)H^2$: At this PM point, the action develops a scalar gauge symmetry
\begin{align}
\label{t0PMspin3symm1}
\delta b_{\mu\nu\rho} &= \nabla_{(\mu}\nabla_\nu\nabla_{\rho)}\chi - \frac{1}{D}g_{(\mu\nu}\nabla_{\rho)}\square\chi+\frac{2(D-1)H^2}{D} g_{(\mu\nu}\nabla_{\rho)}\chi\, ,\\
\delta h & = -\frac{1}{3}\sqrt{2(D-1)H^2}\Big(\square +2(D+1)H^2\Big)\chi.
\label{t0PMspin3symm2}
\end{align}
This removes the helicity-0 polarization, leaving propagating $\pm1$, $\pm 2$, and $\pm3$ helicities.
\end{itemize}
As the mass of a spin-3 field approaches each of these distinguished points, we can define a decoupling limit that makes manifest the branching of the massive (A)dS representation. 
The massless decoupling limit was studied in~\cite{Hinterbichler:2016fgl}, so here we concentrate on the two PM decoupling limits.

\subsubsection*{$t=0$ decoupling limit}
For $s=3$ and $t=0$, the branching rule \eqref{eq:cftbranching} becomes
\be
(\Delta, 3)\xrightarrow[{\Delta\to d-1 }]{}   (d-1,3)\oplus (d+2,0),
\ee
which corresponds to a massive spin-3 field splitting into a $t=0$ PM spin-3 field and a scalar with mass
\be \label{eq:appEmass1}
m^2 = -2(D+1)H^2.
\ee
We now verify this from the free Lagrangian. As we approach the mass $m^2 = 2(D-1)H^2$, the scalar mode is decoupling, so we need to introduce a scalar St\"uckelberg field.
The St\"uckelberg field is introduced by field replacements patterned after the gauge symmetries~\eqref{t0PMspin3symm1} and~\eqref{t0PMspin3symm2},
\begin{align}
\label{t0PMspin3stuck1}
b_{\mu\nu\rho} &\mapsto b_{\mu\nu\rho}+\frac{1}{\epsilon\sqrt{2H^4(D^2-1)}} \left(\nabla_{(\mu}\nabla_\nu\nabla_{\rho)}\phi - \frac{1}{D}g_{(\mu\nu}\nabla_{\rho)}\square\phi+\frac{2(D-1)H^2}{D} g_{(\mu\nu}\nabla_{\rho)}\phi\right),\\
h &\mapsto h -\frac{1}{3\epsilon\sqrt{H^2(D+1)}}\Big(\square +2(D+1)H^2\Big)\phi,
\label{t0PMspin3stuck2}
\end{align}
where the normalization has been chosen to canonically normalize $\phi$ and $\epsilon$ parametrizes the deviation from the PM mass,
\be
\epsilon^2 \equiv m^2 - 2(D-1)H^2.
\ee
After substituting the St\"uckelberg replacements~\eqref{t0PMspin3stuck1},~\eqref{t0PMspin3stuck2}, and taking the $\epsilon \to 0$ limit, the Lagrangian splits into the Lagrangians for a $t=0$ PM field and a scalar,
\be
{\cal L}^{s=3}_{m^2}\xrightarrow[{m^2\to2(D-1)H^2 }]{}  {\cal L}^{s=3}_{ 2(D-1)H^2} - \frac{1}{2}(\partial\phi)^2 +(D+1)H^2\phi^2.
\ee
The scalar field has the mass~\eqref{eq:appEmass1}, precisely as expected.

\subsubsection*{$t=1$ decoupling limit}
We now repeat the above exercise for the $t=1$ decoupling limit. 
The branching rule \eqref{eq:cftbranching} for this case is
\be
(\Delta, 3)\xrightarrow[{\Delta\to d}]{}  (d,3)\oplus (d+2,1),
\ee
which corresponds to a massive spin-3 field splitting into a $t=1$ PM spin-3 field and a vector with mass
\be \label{eq:appEmass2}
m^2 = -3DH^2 .
\ee
To see this from the Lagrangian, we introduce a vector St\"uckelberg field as in~\eqref{t1PMgauge1} and \eqref{t1PMgauge2},
\begin{align}
\label{t1PMstuck1}
 b_{\mu\nu\rho} &\mapsto b_{\mu\nu\rho}+ \frac{1}{\epsilon \sqrt{H^2(D+1)/3}}\left(\nabla_{(\mu}\nabla_{\nu}A_{\rho)} - \frac{1}{D}g_{(\mu\nu}\nabla_{\rho)}\nabla_\alpha A^\alpha+ H^2 g_{(\mu\nu} A_{\rho)}\right)\, ,\\
h &\mapsto h -\frac{1}{\epsilon}\sqrt{\frac{D}{3(D+1)}} \nabla_\mu A^\mu \, ,
\label{t1PMstuck2}
\end{align}
where $\epsilon$ measures the distance from the PM point
\be
\epsilon^2 \equiv m^2 - DH^2.
\ee
Substituting the  St\"uckelberg replacements and taking the decoupling limit $\epsilon\to0$, the free Lagrangian splits into a $t=1$ PM field and a vector,
\be
{\cal L}^{s=3}_{m^2}\xrightarrow[{m^2\to DH^2}]{}  {\cal L}^{s=3}_{DH^2} - \frac{1}{4}F_{\mu\nu}^2 +\frac{3DH^2}{2}A_\mu A^\mu.
\ee
The vector field has the mass~\eqref{eq:appEmass2}, precisely as expected.

\renewcommand{\em}{}
\bibliographystyle{utphys}
\addcontentsline{toc}{section}{References}
\bibliography{shift_arxiv_v3}

\end{document}